\documentclass[floatfix, twocolumn,tighten]{aastex63}
\usepackage[utf8]{inputenc}
\usepackage{amsmath}
\usepackage{amsmath}
\usepackage{pgfplots}
\usepackage{amsmath}
\usepackage{multirow}
\usepackage{amsmath}
\usepackage[maxfloats=256]{morefloats}
\usepackage{hyperref}
\usepackage{tikz}
\usepackage{gensymb}
\pgfplotsset{compat=1.17}
\usepackage{makecell}
\submitjournal{ApJ}
\received{}
\revised{}
\accepted{}

\usepackage{pifont}
\newcommand{\cmark}{\ding{51}}%
\newcommand{\xmark}{\ding{55}}%
\usepackage{array}
\newcolumntype{P}[1]{>{\centering\arraybackslash}p{#1}}

\begin{document}

\title{The nascent milliquasar VT J154843.06+220812.6: tidal disruption event or extreme accretion-state change?}

\author[0000-0001-8426-5732]{Jean J. Somalwar}
\affil{Cahill Center for Astronomy and Astrophysics, MC\,249-17 California Institute of Technology, Pasadena CA 91125, USA.}

\author[0000-0002-7252-5485]{Vikram Ravi}
\affil{Cahill Center for Astronomy and Astrophysics, MC\,249-17 California Institute of Technology, Pasadena CA 91125, USA.}

\author[0000-0001-9584-2531]{Dillon Dong}
\affil{Cahill Center for Astronomy and Astrophysics, MC\,249-17 California Institute of Technology, Pasadena CA 91125, USA.}

\author{Matthew Graham}
\affil{Cahill Center for Astronomy and Astrophysics, MC\,249-17 California Institute of Technology, Pasadena CA 91125, USA.}

\author[0000-0002-7083-4049]{Gregg Hallinan}
\affil{Cahill Center for Astronomy and Astrophysics, MC\,249-17 California Institute of Technology, Pasadena CA 91125, USA.}

\author[0000-0002-4119-9963]{Casey Law}
\affil{Cahill Center for Astronomy and Astrophysics, MC\,249-17 California Institute of Technology, Pasadena CA 91125, USA.}

\author[0000-0002-1568-7461]{Wenbin Lu}
\affil{Department of Astrophysical Sciences, 4 Ivy Lane, Princeton University, Princeton, NJ 08544, USA}

\author{Steven T. Myers}
\affil{National Radio Astronomy Observatory, P.O. Box O, Socorro, NM 87801, USA}

\correspondingauthor{Jean J. Somalwar}
\email{jsomalwar@astro.caltech.edu}

\begin{abstract}
    We present detailed multiwavelength follow up of a nuclear radio flare, VT J154843.06+220812.6, hereafter VT J1548. VT J1548 was selected as a ${\sim}1$ mJy radio flare in 3 GHz observations from the VLA Sky Survey (VLASS). It is located in the nucleus of a low mass ($\log M_{\rm BH}/M_\odot \sim6$) host galaxy with weak or no past AGN activity. VT J1548 is associated with a slow rising (multiple year), bright mid IR flare in the WISE survey, peaking at ${\sim}10\%L_{\rm edd.}$. No associated optical transient is detected, although we cannot rule out a short, early optical flare given the limited data available. Constant late time (${\sim}3$ years post-flare) X-ray emission is detected at ${\sim}10^{42}$ erg s$^{-1}$. The radio SED is consistent with synchrotron emission from an outflow incident on an asymmetric medium. A follow-up, optical spectrum shows transient, bright, high-ionization coronal line emission ($[{\rm Fe\,X}]\,\lambda 6375,[{\rm Fe\,XI}]\,\lambda 7894,[{\rm S\,XII}]\,\lambda 7612$). Transient broad H$\alpha$ is also detected but without corresponding broad H$\beta$ emission, suggesting high nuclear extinction. We interpret this event as either a tidal disruption event or an extreme flare of an active galactic nucleus, in both cases obscured by a dusty torus. Although these individual properties have been observed in previous transients, the combination is unprecedented. This event highlights the importance of searches across all wave bands for assembling a sample of nuclear flares that spans the range of observable properties and possible triggers.
\end{abstract}

\section{Introduction}

Supermassive black holes (SMBHs) at the centers of galaxies power myriad observable phenomena across cosmic time. The evolution of galaxies is closely linked to SMBH activity \citep[e.g.][]{Kormendy2013}. Active galactic nuclei (AGN), which have actively accreting SMBHs at their centers, produce bright multiwavelength emission due to the presence of an accretion disk and, in many cases, an associated jet or outflow \citep[][]{Netzer2015}.

Quiescent or only weakly accreting SMBHs are challenging to study because of their dim or nonexistent emission. The recent advent of high cadence photometric and spectroscopic surveys has enabled the discovery of large samples of tidal disruption events (TDEs), which occur when a star is disrupted as it enters the tidal radius of an SMBH, given by $R_T \sim R_* (M_{\rm BH}/M_*)^{1/3}$ for a black hole of mass $M_{\rm BH}$ and a star of radius (mass) $R_*(M_*)$ \citep[e.g.][]{Rees1976, Rees1988,vanVelzen2011,vanVelzen2019,Donley2002,vanVelzen2021, Sazonov2021}. TDEs provide a key probe of the SMBHs and nuclear regions in quiescent galaxies: among many insights, they enable measurements of the dust covering factors in quiescent galaxies, the circum-nuclear density profile, and they may provide a new method of measuring the mass of low mass (${\sim}10^6\,M_\odot$) SMBHs \citep[e.g.][]{Jiang2021_IRTDE, vanVelzen2019, Mockler2019, Metzger2012}. They are often observed as $10^{41-45}$ erg s$^{-1}$ X-ray transients, which decay with the mass fallback rate as a $t^{-5/3}$ power law \citep[e.g.][]{Bade1996, Komossa1999, Esquej2007}. The X-rays may originate directly from an accretion disk or via material forced inward at the nozzle shock close to pericenter \citep[e.g.][]{Komossa1999,Piran2015,Auchettl2017,Krolik2016}. 

While the landscape of TDEs in quiescent galaxies is rapidly being mapped out, the evolution of a TDE in a galaxy with a pre-existing accretion disk is poorly understood \citep[although, recent simulations are gaining ground, see][]{Chan2020}. Given current knowledge, it is difficult, or in some cases impossible, to observationally differentiate between a nuclear flare caused by a TDE and one caused by an accretion-state change \citep[see][for a review of possible distinguishing characteristics]{Zabludoff2021}. This problem is made particularly challenging because of the many remaining mysteries in accretion disk physics: the magnitude of possible state changes due to accretion disk instabilities, their occurrence rate, and their multiwavelength properties are largely unknown \citep[see][and references therein]{Lawrence2018}.

Thus, nuclear flares from galaxies with pre-existing accretion disks are particularly challenging to interpret. In galaxies where a pre-existing accretion disk cannot be ruled out (i.e., those that are either weakly accreting or are quiescent but were accreting in the recent past), several aspects of the central SMBH and the inner few parsecs of the galaxy remain mysterious. For example, it is still not understood if, when, and how a dusty torus can form in a weakly accreting or non-accreting galaxy \citep[e.g.][]{Honig2007, Hopkins2012}.

Progress in observationally mapping out the range and properties of nuclear flares from weakly accreting or recently accreting galaxies is advancing. For example, searches for transient line emission in the Sloan Digital Sky Survey (SDSS) spectroscopic survey \citep[][]{Strauss2002} have unveiled a class dubbed the extreme coronal line emitters (ECLEs), which show bright, high ionization (${\gtrsim}100$ eV) coronal emission lines (e.g., $[{\rm Fe\,X}]\,\lambda 6375$, $[{\rm Fe\,XIV}]\,\lambda 5303$) \citep[e.g.][]{Komossa2008}. These lines are excited by a transient, high-energy, photoionizing continuum and fade on ${\sim}3{-}5$ yr timescales \citep{Yang2013}. 

Although most of the ${\sim}20$ known ECLEs are in quiescent galaxies \citep[e.g.][]{Wang2011, Wang2012, Frederick2019, Malyali2021,Komossa2008}, an increasingly large subset are hosted by galaxies which lie in the grey area between strongly accreting AGN and quiescent galaxies. For example, ASASSN-18jd was a nuclear transient in a host galaxy with no clear evidence for AGN activity \citep[][]{Neustadt2020}. Although this event had a TDE-like blue continuum and a high ratio of $[{\rm Fe\,X}]$ to $[{\rm O\,III}]$, it showed a non-monotonically declining optical light curve and a harder-while-fading X-ray spectrum that are both more typical of AGN activity. Likewise, the transient AT 2019avd showed strong coronal line emission alongside TDE-like transient features (e.g., soft X-ray emission, Bowen fluorescence lines, broad Balmer emission), and is located in an inactive galaxy \citep[][]{Malyali2021}. Its double peaked optical light curve is characteristic of AGN activity, although some exotic TDE models could predict similar behavior \citep{Malyali2021}.

Originally, ECLEs were thought to be associated with TDEs, which can produce the requisite high energy continuum that would only illuminate the coronal line emitting region but not excite [O\,III] immediately  because of light travel time effects \citep[][]{Wang2012}. However, it is well known that AGN-like continua can produce coronal line emission since, before the discovery of ECLEs, coronal lines were most often observed from Seyfert galaxies of all types \citep[e.g.][]{Seyfert1943, Peterson1984, Gelbord2009}. ${\sim}2/3$ of AGN across the range of activity levels show at least one coronal line in the near-infrared (NIR) \citep[][]{Riffel2006}. This fraction is poorly constrained in the optical because optical coronal lines are dim in most AGN, with the brightest [Fe\,VII]$\lambda6086$ lines no more than ${\sim}10\%$ of the [O\,III]$\lambda$5007 flux \citep[][]{Murayam1998}. An accretion state change could well replicate the ECLE phenomena.

Key evidence in understanding the possible triggers of ECLEs lies in their {\it multiwavelength} emission. ECLEs sometimes show transient, broad lines (FWHM${\sim}1000{-}2000$ km s$^{-1}$), including hydrogen Balmer emission \citep[e.g.][]{Wang2012}. ECLEs have been associated with optical/UV flares, which begin before the coronal lines appear \citep[][]{Palaversa2016, Frederick2019}. Many ECLEs have been associated with IR flares with luminosities ${\sim}10^{42-43}$ erg s$^{-1}$, consistent with emission from dust \citep[e.g.][]{Dou2016}. The IR emission can fade on timescales as long as ${\sim}10$ years \citep[e.g.][]{Dou2016}. The radio emission from ECLEs, which can constrain the presence of a nascent jet or outflow, is practically unconstrained. Note that the relative frequency of the different multi-wavelength signatures in galaxies that may have pre-existing accretion disks and those that are quiescent is unknown.

More conclusive constraints on the trigger(s) of ECLEs require a large sample of events with minimal selection biases. Searches based on evolving optical spectral features may miss objects similar to the ECLEs but with dimmer coronal line emission. The multi-wavelength, transient emission from ECLEs will allow us to understand the full range of possible triggers and host properties.

In this work, we present the first radio selected ECLE, SDSS J154843.06+220812.6, hereafter SDSS J1548. SDSS J1548 shows weak or no evidence for accretion, so a pre-existing accretion disk cannot be ruled out. SDSS J1548 was identified by \cite{Jiang2021} as the host of a bright nuclear MIR flare. Independently, we selected SDSS J1548 as part of our ongoing effort to compile a sample of radio-selected TDE candidates using the VLA Sky Survey (VLASS; \citealp{Lacy2020}). We performed an extensive follow up campaign, during which we identified this object as an ECLE with additional broad Balmer features. It is X-ray bright, although, intriguingly, it shows no optical flare in the available data. The transient emission appears to evolve on long (${\sim}$year) timescales.

We present multi-wavelength observations of SDSS J1548 and the associated transient, which we label VT J1548+2208 (VT J1548 hereafter). In Section~\ref{sec:target_selection}, we describe our target selection. In Section~\ref{sec:obs_red}, we detail both the archival and follow-up observations and data reduction. In Section~\ref{sec:host}, we describe the non-transient galactic-scale properties of SDSS J1548. In Sections~\ref{sec:transientspec} and \ref{sec:transientbroad} we discuss the transient emission associated with VT J1548. Finally, in Section~\ref{sec:discussion} we consider the possible origins (i.e., TDE, AGN-related activity) of VT J1548, and in Section~\ref{sec:conclusion} we conclude.

We adopt a standard flat $\Lambda$CDM model with H$_0 = 70$ km s$^{-1}$ Mpc$^{-1}$ and $\Omega_{\rm m} = 0.3$. All magnitudes are reported in the AB system unless otherwise specified.

\section{Target Selection} \label{sec:target_selection}

We selected VT J1548 during our search for radio-bright TDE candidates using the Karl G. Jansky Very Large Array (VLA) Sky Survey (VLASS). VLASS is a full-sky, radio survey ($\delta > -40^\circ$, $2-4$ GHz; \citealp{Lacy2020}). Each VLASS pointing will be observed three times. The first epoch (E1) was completed between $2017{-}2018$ and the second (E2) is halfway done (${\sim}2020{-}$present). VLASS is optimal for studies of radio-emitting TDEs because it is sensitive (${\sim}0.13\,$mJy) and has a high angular resolution that allows for source localization to galactic nuclei (${\sim}2\farcs5$, with variations with declination and hour angle).

Dong et al{.}, in prep{.}, developed a pipeline to robustly identify radio transients with VLASS, which we used to select radio TDE candidates. We will describe the source detection and photometry in detail in that work; we provide a brief summary in Appendix~\ref{sec:VLASStrans}. 

We selected TDE candidates as nuclear VLASS transients (${<}3\arcsec$ from the center of a Pan-STARRS source; \citealp{Flewelling2020, Chambers2016}) with no archival radio detections (${>}3\arcsec$ from a source in the NVSS or FIRST catalogues; \citealp{Condon1998, Helfand2015, White1997}). After this initial selection, we verified that each source was nuclear using precise positions from VLA follow up. We required the stellar mass of the host galaxy, measured using an SED fit (Section~\ref{sec:host}), to be consistent with $\log M_{\rm BH}/M_\odot \lesssim 8$  according to the stellar mass - SMBH mass relation from \cite{Greene2020} (i.e., $\log M_*/M_\odot \lesssim 12$). For SMBH masses $\log M_{\rm BH}/M_\odot \gtrsim 8$, stars will be captured whole rather than be disrupted because the Hill radius is comparable to the tidal radius  \citep[][]{Rees1988}. After this initial selection, we carefully inspected the archival radio images to ensure there are no sub-threshold detections. We will present the full sample of radio selected TDE candidates in future papers. In this paper, and other in prep{.} work, we present individual, unique candidates, including VT J1548.

\section{Observations and Data Reduction} \label{sec:obs_red}

After identifying VT J1548 as a promising TDE candidate, we performed extensive, multi-wavelength follow up. In this section, we describe the observations and data reduction. We also present the available archival data. Detailed data analysis and interpretation will be described in later sections. Figure~\ref{fig:summary} summarizes the observation timeline. 

\subsection{Radio Observations} \label{sec:rad_red}

SDSS J1548 was undetected in the NVSS and FIRST radio surveys \citep{Condon1998, Helfand2015, White1997}. Most recently, it was observed on MJD $58046$ (Oct. 15, 2018) during VLASS E1 with a $3\sigma$ upper limit $f_{\nu}({\rm 3\,GHz}) < 0.36$ mJy. VT J1548 was first detected in the radio during VLASS E2 on MJD $59068$ (Aug. 7, 2020) with $f_{\nu}({\rm 3\,GHz}) = 1.12\pm0.15$ mJy.

We obtained a broadband ($0.3{-}20$ GHz) radio SED for VT J1548 on MJD $59273$ (Feb. 28, 2021) as part of program 20B-393 (PI: Dong). We reduced the data using the Common Astronomy Software Applications (CASA) with standard procedures. VT J1548 was detected in the L, S, C, and X bands and undetected in the P band.

\subsection{Optical/IR Light Curve}

SDSS J1548 is in the survey area of the NEOWISE and Zwicky Transient Facility (ZTF) surveys \citep[][]{Mainzer2011, Bellm2019, Graham2019}. NEOWISE has observed SDSS J1548 in the W1 (3.4 $\mu$m) and W2 (4.6 $\mu$m) bands with a cadence of  ${\sim}6$ months since MJD $\sim 56700$. Each epoch consists of ${\sim}12$ exposures. We downloaded the NEOWISE photometry from \url{irsa.ipac.caltech.edu}. The lightcurve is shown in Figure~\ref{fig:summary}. SDSS J1548 flared brightly in NEOWISE beginning on MJD${\sim}58100$ (Mar. 23 2018). It increased from $W1/W2 \sim 13.6/13.6$ mag (native Vega system) to $W1/W2 \sim 11.2/10.3$ mag (native Vega system) in ${\sim}900$ days and had not begun to fade by the most recent observation (MJD $59049$; Jul. 19 2020). The peak flux of the flare was $\gg 5\sigma_{\rm quies.}$, where $\sigma_{\rm quies.}$ was the root-mean-square variability in the pre-flare NEOWISE data.

ZTF is a high cadence optical transient survey. SDSS J1548 was observed as part of the public MSIP survey \citep{Bellm2019}, which observes the full northern sky every three nights in the $gr$ filters. We used the IPAC forced photometry service \citep[][]{Masci2019} to download the optical light curve, and processed it using the recommended signal-to-noise cuts\footnote{\url{http://web.ipac.caltech.edu/staff/fmasci/ztf/forcedphot.pdf}}. No optical transient is detected in the available data, although we may have missed the transient because of poor coverage. MJD $57500{-}58000$ is only covered by the ATLAS survey, but the ATLAS coverage has a gap between MJD $57650{-}57750$, and it is possible that an optical transient would be undetected if it occurred near $57500$ and contaminated the ATLAS reference images. Assuming no systematic problems in the photometry that may mask a flare, we can exclude an optical transient that peaks during the ATLAS coverage with a flux density brighter than ${\sim}0.6$ mJy ($L\lesssim 6\times10^{42}$ erg s$^{-1}$) at the $5\sigma$ level in the ATLAS $o$ band. This constraint rules out a flare similar to those in optically-selected TDEs \citep{vanVelzen2021}, unless it occurred between MJD $57650{-}57750$.

\subsection{X-ray Observations} \label{sec:xray_red}

SDSS J1548 is not detected in any archival X-ray catalogs, including the Second ROSAT All-Sky Survey Point Source Catalog \citep[][]{Voges1993, Boller2016}. The best limit on the host galaxy X-ray flux is from a serendipitous 17.9 ks {\it XMM-Newton} exposure ${\sim}100$ days before the first VLASS epoch (PI: Seacrest, MJD 57950; Jul. 16 2017). We retrieved the Processing Pipeline System (PPS) products from the {\it XMM-Newton} archive. The PPS products have already been reduced using standard procedures with the most up-to-date pipeline and calibration files. We used the \texttt{ximage} \texttt{sosta} tool to measure the source flux at the location of SDSS J1548 on the EPIC-PN and MOS2 $0.2-12$ keV images \citep[][]{Giommi1992}. (SDSS J1548 was not in the field-of-view of the EPIC-MOS1 image.) We used the recommended source box size. However, SDSS J1548 is near the edge of both images, so the recommended background box sizes extended off the image. We manually drew background boxes of different sizes centered on/near the source and measured the intensity in each case, to verify that our choice did not affect our result. The source was undetected, with a $3\sigma$ upper limit on the $0.2-10$ keV flux of ${\sim}10^{-13}$ erg cm$^{-2}$ s$^{-1}$. We get a similar upper limit using both the PN and MOS1 images, which suggests that our result is not strongly affected by the fact that the source is near the image edge.

SDSS J1548 was observed three times (MJD 59127/Oct. 5 2020, 59281/Mar. 8 2021, 59388/Jun. 23 2021) post-flare with ${\sim}2$ ks exposures by the Swift X-ray Telescope (Swift/XRT; \citealp{Burrows2005}). The final epoch was a target of opportunity (ToO) observation requested by our group. The first two observations are ToOs (PI Dou) that we found during a search of the Swift archive. The data were reduced using the Swift HEASOFT online reduction pipeline\footnote{\url{https://www.swift.ac.uk/user_objects/index.php}} with default settings to generate a lightcurve at the position of SDSS J1548 \citep[][]{Evans2007}. There is no significant evolution between observations. Hence, we assume no variation in the X-ray emission and stack the observations to obtain a S/N sufficient to extract a spectrum. We used the online Swift pipeline processing implementation of \texttt{xselect} to extract the $0.2-10$ keV spectrum \citep[][]{Evans2009}. The spectrum is shown in Figure~\ref{fig:xray_spec} and we discuss it in Section~\ref{sec:xray}.

\subsection{Optical Spectroscopy}

SDSS J1548 was observed on MJD 53556 (Jul. 5 2005) as part of the SDSS Spectroscopic Survey \citep[][]{Strauss2002}. We retrieved the archival optical spectrum from the SDSS archive. After identifying SDSS J1548 as a transient host, we observed it with the Keck I Low Resolution Image Spectrometer (LRIS; \citealp{Oke1995}) on MJD $59259$ (Feb. 14 2021) and $59348$ (May 14 2021) with exposure times of $10$ and $30$ min{.} respectively. Because of poor seeing, we used the $1\farcs5$ slit for the first epoch, but we used the $1\farcs0$ slit for the second epoch. The slit positions are shown in Figure~\ref{fig:host}. For both epochs, we used the 400/3400 grism, the 400/8500 grating with central wavelength 7830, and the 560 dichroic. This leads to a usable wavelength range of ${\sim}1300{-}10000\,{\rm \AA}$ and a resolution $R{\sim}700$. 

We reduced the first epoch of observations using the \texttt{lpipe}v2020.09 pipeline with default settings \citep[][]{Perley2019}. The LRIS red CCD was upgraded before the second epoch of observations and was incompatible with earlier \texttt{lpipe} versions, so we reduced this deeper epoch using \texttt{lpipe}v2021.06${\rm \beta}$. 

We observed SDSS J1548 on MJD 59371 (Jun. 6 2021) with the Echellette Spectrograph and Imager (ESI; \citealp[][]{Sheinis2002}) on Keck II. ESI is optimal for velocity dispersion measurements because of its resolution, which can be as high as $R\sim13000$ (22.4 km s$^{-1}$ FWHM) in echellette mode. ESI in echellette mode has a wavelength coverage ${\sim}0.4-1.1\,{\rm \mu m}$. We exposed for $25$ minutes using the $0\farcs3$ slit. The slit positioning is shown in Figure~\ref{fig:host}. We reduced the observations using the \texttt{makee} pipeline with the standard star Feige 34. We used default settings, except to adjust the spectral extraction aperture, as described in Appendix~\ref{sec:esi}.

\section{Host Galaxy Analysis} \label{sec:host}

\begin{figure}
    \centering
    \includegraphics[width=0.49\textwidth]{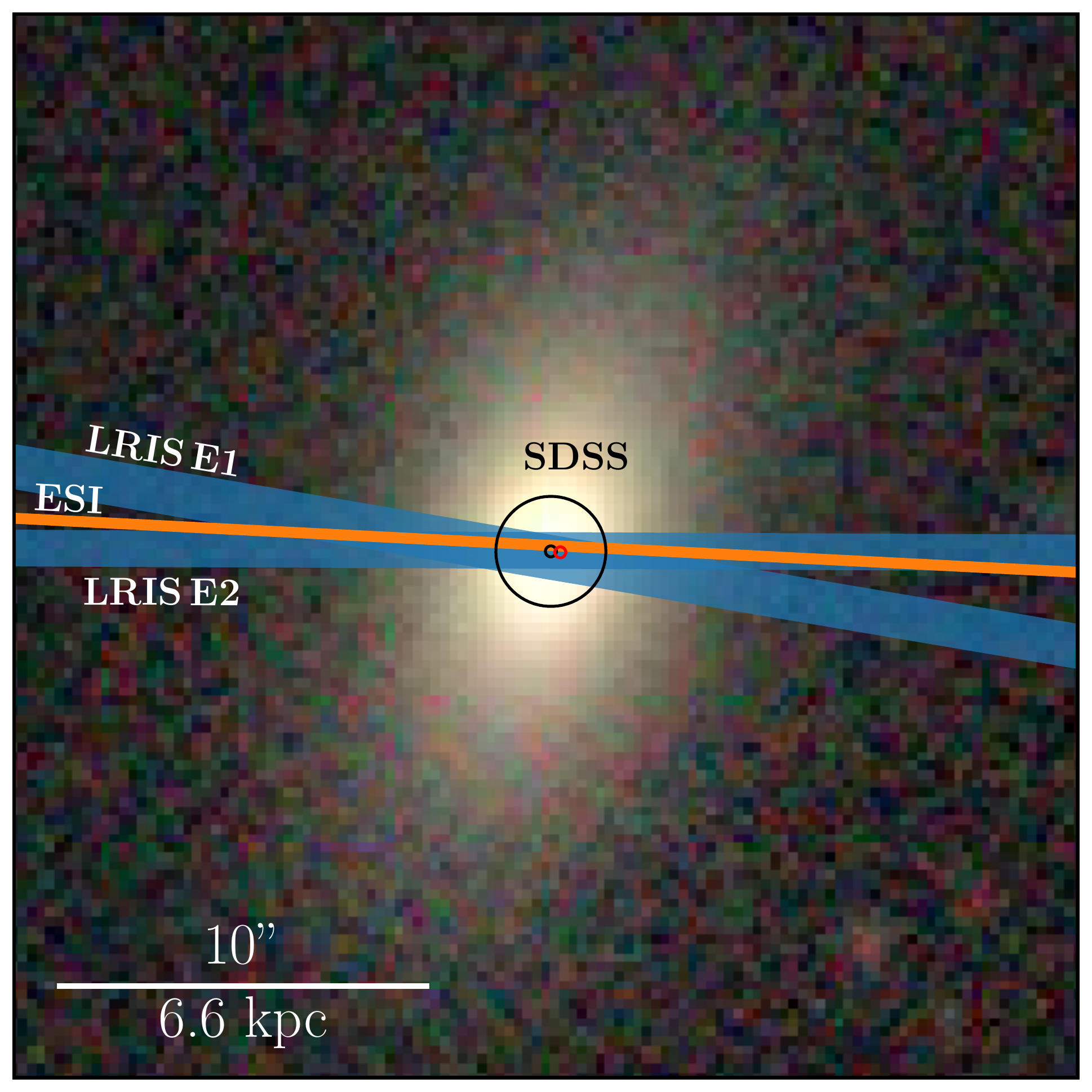}
    \caption{$zrg$ image of SDSS J1548, the host galaxy of the radio transient VT J1548. The optical nucleus is shown as a small red circle and the radio transient position is shown as small black circle. The circle radii show approximate $3\sigma$ statistical uncertainties. Systematic reference-frame uncertainties are not included. The radio transient is consistent with being nuclear. The blue rectangles show the slit positions for LRIS follow up and the orange rectangles shows that for ESI follow up. Image from the Legacy Survey \citep[][]{Dey2019}. }
    \label{fig:host}
\end{figure}

In this section, we describe SDSS J1548, the host galaxy of VT J1548. SDSS J1548 is at redshift $z\sim0.031$ ($d_L \sim 137$ Mpc). Figure~\ref{fig:host} shows a $zrg$ image of SDSS J1548. We have noted the cataloged position of the galaxy nucleus \citep[][]{York2000} and the radio transient position from our VLA follow up. The radio transient is consistent with being nuclear. 

SDSS J1548 is classified as an elliptical or S0 galaxy with a $g$-band semi-major half-light axis ${\sim}1.8$ kpc \citep[]{Huertas-Company2010, Simard2011}. It is bulge-dominated, with a $g$- ($r$-) band bulge-to-disk ratio $B/T = 0.7(0.73)$ \citep[][]{Simard2011}. The bulge-dominated morphology is unusual for ECLEs $-$ the known ECLEs are largely located in intermediate-luminosity disk galaxies with no apparent bulge in SDSS imaging \citep[][]{Wang2012}.

We measured the galaxy stellar mass using an SED fit following \cite{vanVelzen2021} and \cite{Mendel2014}. We retrieved archival photometry from the GALEX (FUV, NUV; \citealp{Million2016, Martin2005}), SDSS ($ugriz$; \citealp{Ahumada2020}), and WISE ($W1$, $W2$; \citealp{Wright2010}) surveys. We used \texttt{prospector} \citep[][]{Johnson2021}, a Bayesian wrapper for the \texttt{fsps} stellar population synthesis tool \citep[][]{Conroy2010, Conroy2009}, with a \cite{Chabrier2003} IMF, a $\tau$-model star formation history, and the \cite{Calzetti2000} attenuation curve. We fixed the redshift to the best-fit redshift from the LRIS spectrum (0.031; Appendix~\ref{sec:spectral_fitting}). We fit the SED using the \texttt{emcee} Monte Carlo Chain Ensemble sampler \citep[][]{Foreman-Mackey2013} with $500\,({\rm burn{-}in})+1000$ steps. The best-fit stellar mass is reported in Table~\ref{tab:host}. Our best-fit parameters are consistent with cataloged SED fits of this source from SDSS. We relate the stellar mass to the SMBH mass using the empirically derived $M_*-M_{\rm BH}$ relation from \cite{Greene2020}. We find $\log M_{\rm BH}\sim 7.1 \pm 0.79$, where the uncertainty is dominated by intrinsic scatter in the relation.

The SMBH mass is more tightly correlated with the bulge velocity dispersion ($\sigma_*$) than $M_*$. We measured $\sigma_*$ from the high resolution ESI spectrum and find an SMBH mass $\log M_{\rm BH}/M_\odot = 6.48 \pm 0.33$, as described in Appendix~\ref{sec:esi}. The error is dominated by intrinsic uncertainty in the $M_{\rm BH}-\sigma_*$ relation. This SMBH mass is consistent with that measured by \cite{Jiang2021} using the lower resolution archival SDSS spectrum. It corresponds to an Eddington luminosity of $3\times10^{44}$ erg s$^{-1}$ \citep[][]{Gezari2021}.

\begin{figure*}
    \centering
    \includegraphics[width=0.99\textwidth]{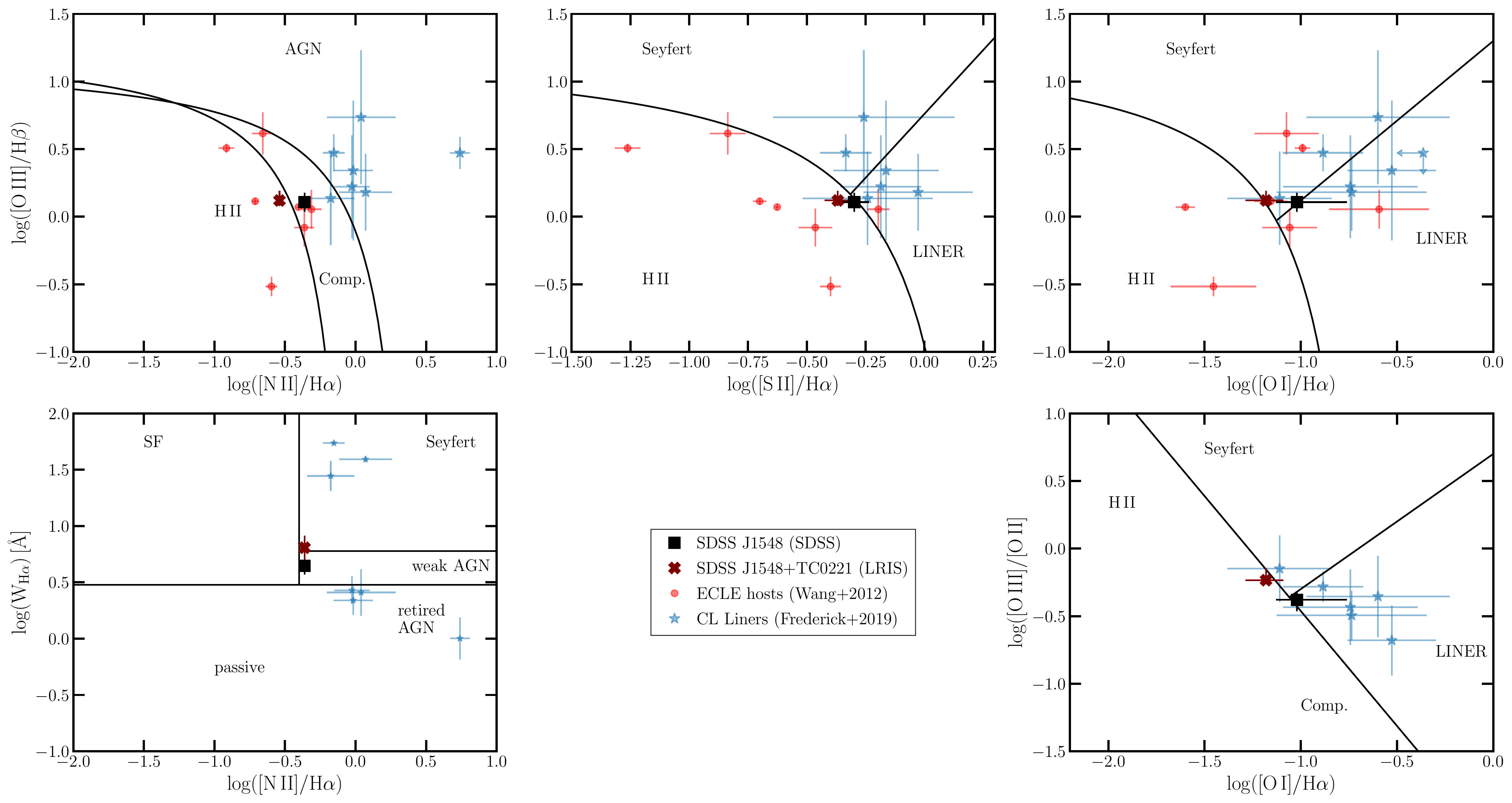}
    \caption{ Five variations of the BPT diagram \citep[][]{Baldwin1981, Kewley2006, Cid2011}, following Figure 13 from \cite{Frederick2019}. The SDSS measurements of SDSS J1548 are shown as a black square, while the LRIS measurements are shown as a reddish cross. Where possible, we include the changing look LINERs from \cite{Frederick2019} and the ECLEs from \cite{Wang2012} for comparison. SDSS J1548 has weak or no AGN activity. }
    \label{fig:BPT}
\end{figure*}

\begin{figure*}[pbth]
    \includegraphics[width=\textwidth]{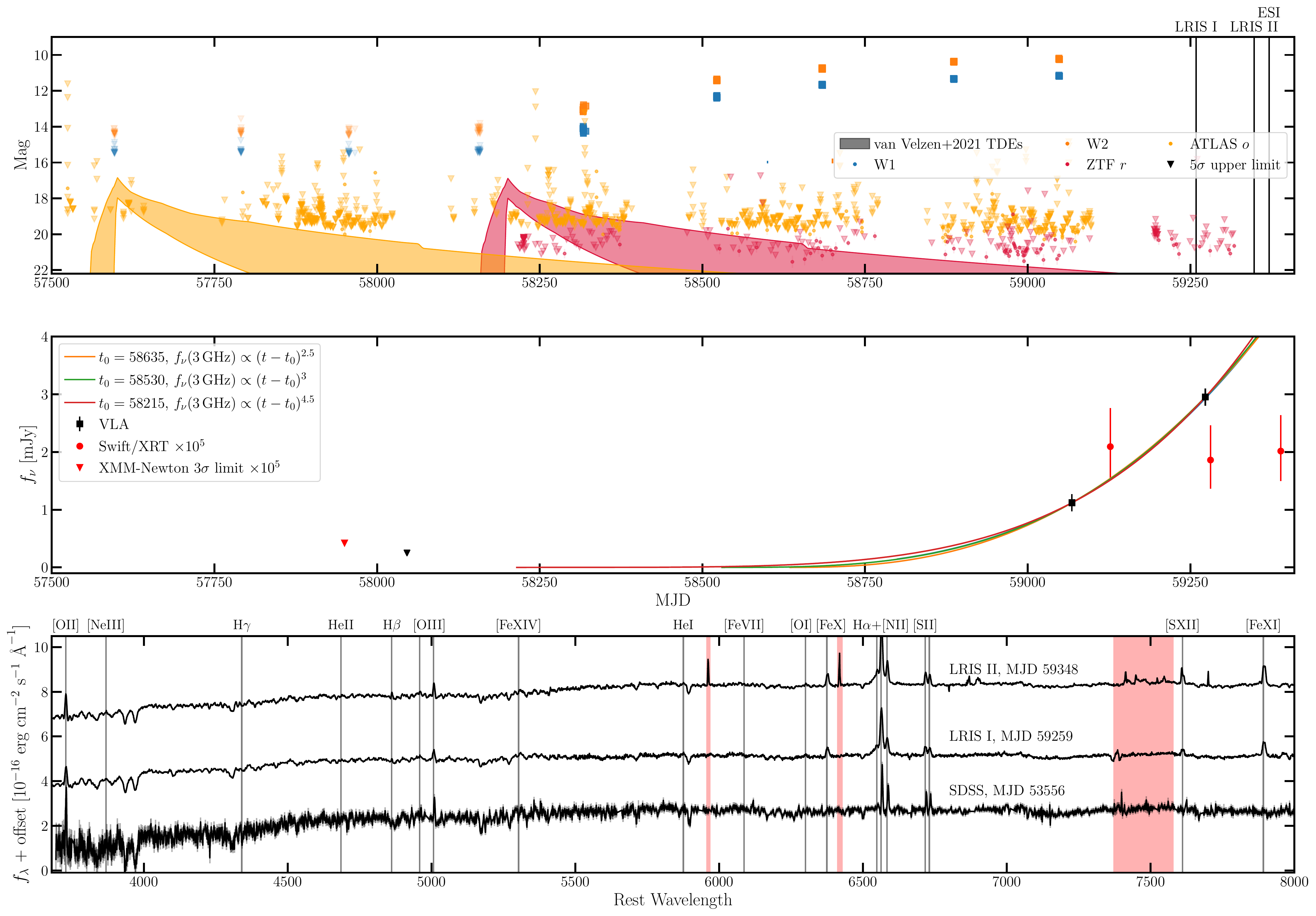}
    \caption{ A summary of the observations of SDSS J1548. ({\it top}) The optical and IR light curves for SDSS J1548. IR magnitudes are in their native Vega system, whereas optical magnitudes are in their native AB system. The IR light curve flares by ${\sim}2-3$ mag beginning around MJD $58100$. There is no obvious optical flare, although gaps in survey coverage mean that we cannot rule one out. The optical upper limits are consistent with the tail of the best-fit blackbody to the IR emission. We overplot the range of fluxes expected in different bands (as denoted by the color of the band) from typical optical TDEs as shaded regions. We have adopted the models of ZTF optical TDE light curves from Table 6 of \cite{vanVelzen2021} with the appropriate distance for SDSS J1548 and extinction $E(B-V)_{\rm nuc} \sim 1$. We plot the central $\pm 1\sigma$ range of fluxes expected from these models. We shift the start date of the each band to be within the coverage of the corresponding survey but still consistent with the start of the WISE flare. ({\it middle}) The 3 GHz radio (black squares) and $0.2-10$ keV X-ray light curves. Solid lines show power law fits to the radio light curve, with the launch date in each case noted in the legend. ({\it bottom}) The optical spectrum evolution. The late-time LRIS optical spectra show transient broad Balmer and coronal line emission. We highlight regions impacted by reduction problems in red. }
    \label{fig:summary}
\end{figure*}

Next, we constrain any prior AGN activity in SDSS J1548. The archival SDSS spectrum is shown in the bottom panel of Figure~\ref{fig:summary}. It has many narrow features, including the Balmer series, $[{\rm O\,III}]\,\lambda 5007$, $[{\rm N\,II}]\,\lambda 6548, 6584$, and $[{\rm S\,II}]\,\lambda 6713, 6731$, but no broad emission. We fit the narrow lines following Appendix~\ref{sec:spectral_fitting} and the fluxes are tabulated in Table~\ref{tab:line_rat}. Figure~\ref{fig:BPT} shows five variations of the BPT diagrams, which classify galaxies according to their AGN activity \citep[][]{Baldwin1981, Kewley2006, Cid2011}. We plot the ECLE hosts from \cite{Wang2012} and changing look (CL) LINERs from \cite{Frederick2019}, where possible. The CL LINER sample includes one ECLE (see discussion in Section~\ref{sec:discussion}). SDSS J1548 lies between the ECLE and CL LINER samples. It is consistent with weak or no AGN emission. 

The WISE color of a galaxy (pre-transient) provides an additional constraint on its AGN activity \citep[][]{Assef2018}. The WISE color W1${-}$W2$=0.055$ (W1/W2 = $13.625\pm0.025/13.570 \pm 0.029$) is inconsistent with typical AGN, which have W1${-}$W2$\gtrsim 0.8$ \citep[][]{Assef2018}. Hence, SDSS J1548 may be quiescent or weakly active. Note that the current NEOWISE color (W1${-}$W2$\sim0.9$) is in the AGN regime.

\begin{deluxetable}{cc}
\tablecaption{Host Galaxy  \label{tab:host}}
\tablewidth{0pt}
\tablehead{\colhead{Parameter} & \colhead{Value} }
\startdata
R.A. & 15{:}48{:}43.06 \\
Dec. & 22{:}08{:}12.84 \\
Redshift $z$ & 0.031 \\
$d_L$ & 137 Mpc \\
$\log M_*/M_\odot$ & $10.15^{+0.07}_{-0.07}$ \\
$\log M_{\rm BH}/M_\odot$ (from $M_{\rm BH}-\sigma_*$) & $6.48 \pm 0.33$
\enddata
\tablecomments{R.A. and Dec. are from the SDSS imaging survey \citep[][]{York2000}. Redshift is as measured in our work. The stellar mass is derived from an SED fit. SMBH mass is measured using the velocity dispersion and the \cite{Kormendy2013} $M_{\rm BH}-\sigma_*$ relation.  }
\end{deluxetable}

Finally, SDSS J1548 is within the virial radius of a small group (total halo mass ${\sim} 10^{11.5}\,M_\odot$; \citealp[][]{Saulder2016}). SDSS J1548 shows no obvious evidence for a disturbed morphology indicative of a recent interaction or merger.

\section{Analysis of transient spectral features} \label{sec:transientspec}

Next, we consider the transient emission associated with VT J1548, summarized in Figure~\ref{fig:summary}. We begin by describing the transient spectral features, which will inform our discussion of the broadband emission in the next section. We identify transient lines as those present in the LRIS spectra but not in the SDSS spectrum.

First, we provide a brief summary of the transient features. The following subsections will analyze specific features in detail. The line fluxes for each observation epoch, measured using the procedure described in Appendix~\ref{sec:spectral_fitting}, are listed in Table~\ref{tab:line_rat}. 

VT J1548 was associated with the appearance of strong, high-ionization coronal line emission. We detect $[{\rm Fe\,X}]\,\lambda 6375$, $[{\rm Fe\,XI}]\,\lambda 7892$, and $[{\rm S\,XII}]\,\lambda 7611$. $[{\rm Fe\,XIV}]\,\lambda 5303$ is marginally detected, and we do not observe any [Fe\,VII] emission. The coronal lines are all double peaked, with two roughly equal flux components separated by ${\sim}230$ km s$^{-1}$. We will discuss these lines in detail in Section~\ref{sec:coronal}. We also detect a broad H$\alpha$ component (FWHM$\sim1900$ km s$^{-1}$), although we do not detect the corresponding H$\beta$ feature (Section~\ref{sec:broad}). We refrain from a detailed line flux evolution analysis because the flux calibration may be imperfect. The transient line fluxes agree within $2\sigma$ between the LRIS observations. The lines which are present in all observations (SDSS+LRIS) do not evolve between epochs, except that the narrow H$\alpha$ brightens. This brightening could be caused by the changing slit widths if the H$\alpha$ is extended, so we do not consider it further. 

${\rm He\,II}\,\lambda 4686$ is commonly observed in TDE candidates accompanied by N\,III\,$\lambda4640$ due to the Bowen fluorescence mechanism \citep[][]{Gezari2021}. We do not observe these lines. [Fe\,II] lines are abundant in Seyferts but are undetected from VT J1548 \citep[][]{Mullaney2008}.

\subsection{Coronal line emission} \label{sec:coronal}

The strongest observed coronal lines are $[{\rm Fe\,X}]\,\lambda 6375$ (ionization potential 262.1 eV), $[{\rm Fe\,XI}]\,\lambda 7894$ (IP 290.9 eV), and $[{\rm S\,XII}]\,\lambda 7612$ (IP 564.41 eV) with luminosities $(1.2, 2.3, 1.3)\times10^{39}$ erg s$^{-1}$, respectively (we have summed over all velocity components, see discussion later in this section). The ${\rm [O\,III]}\,\lambda5007$ luminosity is ${\sim}1.2\times10^{39}$ erg s$^{-1}$. The [Fe\,X] to [O\,III] ratio of ${\sim}1$ is unprecedented for ``standard'' Seyferts, which typically have coronal line luminosities that are a factor of ${\sim}100$ dimmer than [O\,III] (see Figure 5 of \citealp{Wang2012}). These fluxes are also marginally dimmer than observed in other ECLEs, which have $L_{\rm Fe\,X,\,lit.} \gtrsim 3\times10^{39}$ erg s$^{-1}$ despite similarly low SMBH masses \citep{Wang2012}. Selection effects may explain the brighter coronal lines in many ECLEs. Alternatively, VT J1548 may be more obscured than the \cite{Wang2012} ECLEs.

We marginally detect $[{\rm Fe\,XIV}]\,\lambda 5303$ at ${<}2\sigma$ significance. Most ECLEs with $[{\rm Fe\,XIV}]\,\lambda 5303$ emission have $L_{\rm [Fe\,XIV]} \gtrsim 0.1 L_{\rm [Fe\,X]}$ \citep[][]{Wang2012}. We expect sufficiently high energy photons to ionize [Fe\,XIV] because it has a lower ionization potential than the bright $[{\rm S\,XII}]$ line. Extinction could weaken the [Fe\,XIV] emission: [Fe\,XIV] is the bluest of the coronal lines. If the coronal lines are heavily extincted, like the broad Balmer emission (see next section), the [Fe\,XIV] line could be extincted by a factor of ${\sim}1.5-2$ relative to [Fe\,X]. This extinction is unlikely to affect the ECLE classification because reducing the [Fe\,X] to [O\,III] ratio by a factor of ten would require $E(B-V)\gtrsim 5$.

\begin{deluxetable*}{c|ccc}
\centerwidetable
\tablecaption{Optical Emission Line Strengths  \label{tab:line_rat}}
\tablewidth{10pt}
\tablehead{\colhead{Line} & \colhead{SDSS}  & \colhead{LRIS I}  & \colhead{LRIS II} }
\startdata
H$\alpha$ (narrow) & $15.60_{-0.35}^{+0.24}$ & $16.95_{-2.78}^{+2.54}$ & $23.34_{-0.15}^{+0.09}$ \\
H$\alpha$ ($2060$ km s$^{-1}$) & $-$ & $18.09_{-4.22}^{+3.84}$ & $26.32_{-1.48}^{+1.37}$ \\
H$\beta$ & $4.69_{-0.61}^{+0.61}$ & $3.88_{-0.39}^{+0.50}$ & $3.84_{-0.31}^{+0.35}$ \\
${\rm [O\,I]}\,\lambda 6300$ & $1.49_{-0.37}^{+0.89}$ & $1.41_{-0.23}^{+0.27}$ & $1.54_{-0.38}^{+0.32}$ \\
${\rm [Fe\,X]}\,\lambda 6375$ & $-$ & $2.31_{-1.36}^{+1.10}/2.73_{-1.71}^{+0.64}$ & $3.18_{-1.01}^{+0.70}/2.56_{-1.97}^{+0.62}$ \\
${\rm [Fe\,XI]}\,\lambda 7894$ & $-$ & $3.94_{-0.62}^{+0.51}/4.60_{-0.52}^{+0.40}$ & $5.37_{-0.44}^{+0.26}/5.50_{-0.46}^{+0.29}$ \\
${\rm [Fe\,XIV]}\,\lambda 5303$ & $-$ & $0.65_{-0.42}^{+3.31}$ & $0.50_{-0.30}^{+0.42}$ \\
${\rm [S\,XII]}\,\lambda 7612$ & $-$ & $1.73_{-0.32}^{+0.31}/1.55_{-0.42}^{+0.26}$ & $2.10_{-0.23}^{+0.19}/3.95_{-0.25}^{+0.25}$ \\
${\rm [O\,II]}\,\lambda\lambda 3726,3729$ & $14.40_{-2.34}^{+1.74}$ & $10.23_{-0.41}^{+0.59}$ & $8.67_{-0.43}^{+0.36}$ \\
${\rm [O\,III]}\,\lambda 4959$ & $2.35_{-0.44}^{+0.69}$ & $1.82_{-0.49}^{+0.31}$ & $1.66_{-0.24}^{+0.31}$ \\
${\rm [O\,III]}\,\lambda 5007$ & $6.01_{-0.62}^{+0.54}$ & $5.69_{-0.25}^{+0.48}$ & $5.06_{-0.25}^{+0.25}$ \\
${\rm [N\,II]}\,\lambda 6548$ & $2.15_{-0.54}^{+0.28}$ & $1.82_{-0.73}^{+0.79}$ & $1.34_{-0.43}^{+0.48}$ \\
${\rm [N\,II]}\,\lambda 6584$ & $6.79_{-0.44}^{+0.37}$ & $6.57_{-0.56}^{+0.48}$ & $6.75_{-0.39}^{+0.42}$ \\
${\rm [S\,II]}\,\lambda 6716$ & $3.84_{-0.53}^{+0.90}$ & $4.55_{-0.83}^{+0.82}$ & $5.39_{-1.22}^{+0.43}$ \\
${\rm [S\,II]}\,\lambda 6731$ & $4.01_{-0.74}^{+0.69}$ & $4.51_{-0.80}^{+0.81}$ & $4.61_{-0.38}^{+1.21}$ \\
\enddata
\centering
\tablecomments{Line fluxes are reported in units of $10^{-16}$ erg cm$^{-2}$ s$^{-1}$. While we report absolute fluxes, the flux calibration is likely imperfect.}
\end{deluxetable*}

We do not detect [Fe\,VII] emission although it has a low ionization potential \citep[][]{Wang2012}. There are a number of ECLEs with undetected [Fe\,VII], and most have been attributed to TDEs \citep{Wang2012}. These ECLEs tend to be galaxies that are less luminous and lower mass than those with detected [Fe\,VII], which is consistent with the low SMBH mass measured for SDSS J1548 \citep{Wang2012}. Moreover, if they are associated with an optical flare, the flare is dimmer than in those galaxies with [Fe\,VII] detections \citep[][]{Wang2012}. The low statistics in current ECLE samples render these trends inconclusive.

\cite{Wang2012} suggest that [Fe\,VII] dim ECLEs can be explained if either the [Fe\,VII] is collisionally de-excited because of its low critical density ($10^{6-7}$ cm$^{-3}$ compared to $>10^{9}$ cm$^{-3}$ for the higher ionization iron lines), or if the X-ray SED is sufficiently bright and peaked above ${\sim}250$ eV so that higher ionization states are favored. The first scenario is disfavored if coronal line emission from ECLEs is produced analogously to that in Seyfert galaxies. In Seyferts, [Fe\,VII] is expected to be emitted from gas which is lower density and more extended than that which emits the higher ionization Fe lines. For example, \cite{Gelbord2009} suggest that the coronal line-emitting gas is embedded in a wind, and the [Fe\,VII] emitting gas is upstream of the gas which emits the higher ionization Fe lines. If this model also applies to ECLEs, it is unlikely that {\it all} of the coronal line-emitting gas is above the [Fe\,VII] critical density.

An excess of soft photons can cause a high [Fe\,X]/[Fe\,VII] ratio. \cite{Gelbord2009} discuss a few Seyferts with high [Fe\,X]/[Fe\,VII] which also have high [Fe\,X]/[O\,III] ratios (although not as extreme as ECLEs) and broad H$\alpha$ FWHM which are narrower than expected (${\sim}750$ km s$^{-1}$). They argue that these extreme ratios are related to the X-ray SED shape. A soft excess which drops off around 100 eV would cause [Fe\,X]/[Fe\,VII] to be high, although it is unclear whether this would explain the extreme ratios observed in [Fe\,VII] dim ECLEs. Alternatively, the soft excess can continue below 100 eV if the [Fe\,VII] emitting gas is obscured from the photoionizing source. As \cite{Wang2012} discusses in the context of ECLEs, a very bright soft X-ray source that overionizes the coronal line-emitting gas could also explain the [Fe\,VII] non-detections. 

\begin{figure*}
    \centering
    \includegraphics[width=\textwidth]{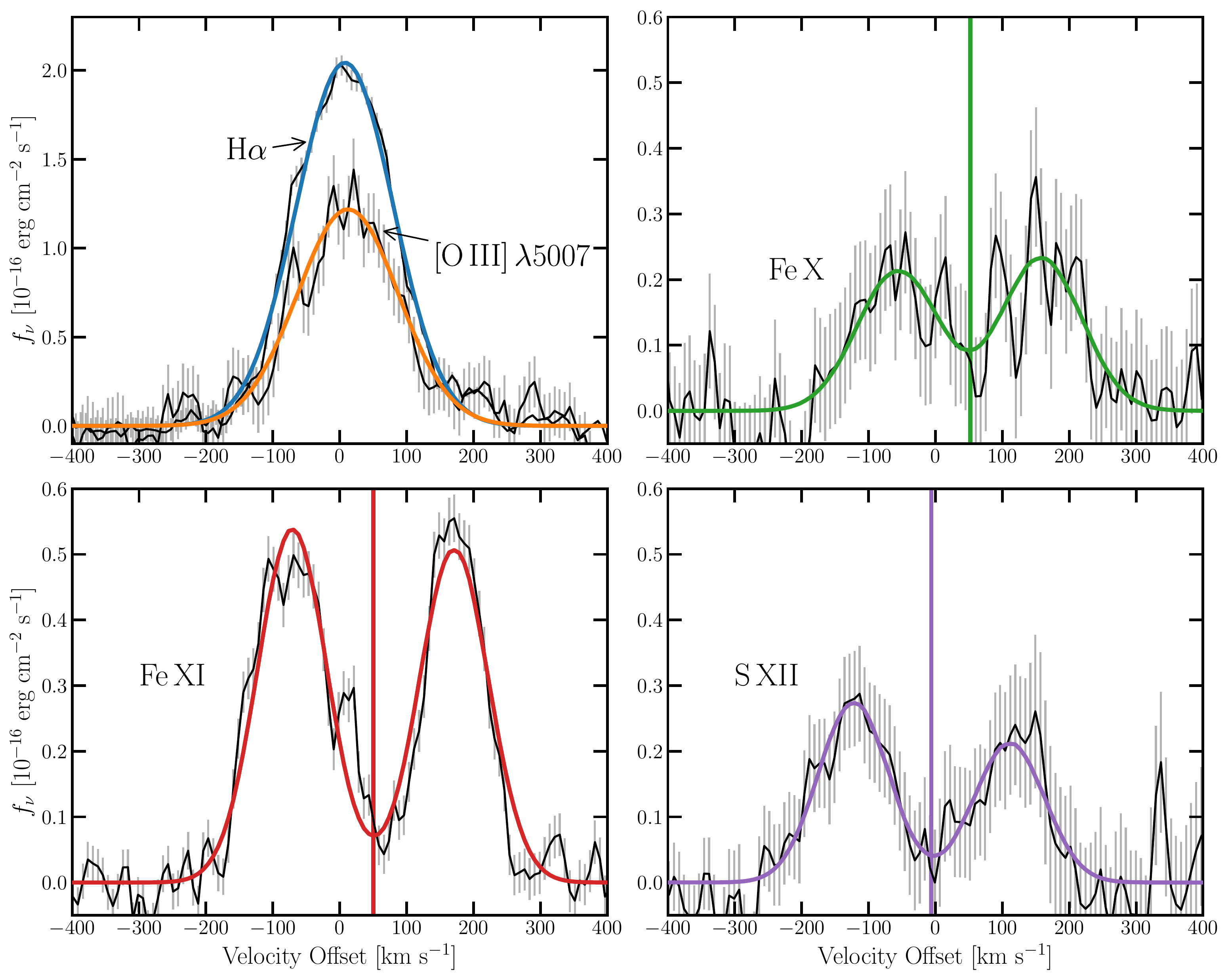}
    \caption{Line profiles for selected lines from the ESI observations of SDSS J1548. The {\it top left} panel shows the H$\alpha$ and [O\,III] line profiles. The faint lines show the observations and the solid lines show Gaussian fits, where we include two Gaussian components in each case to match the coronal line profiles. The {\it top right} and {\it bottom} panels show the coronal line profiles, which all clearly contain two Gaussian components separated by $215-240$ km s$^{-1}$. The solid lines in between each pair of lines indicates the average of the two peak wavelengths.}
    \label{fig:profiles}
\end{figure*}

Further insight into the origin of the coronal line emission comes from close inspection of the coronal line profiles in the high resolution ESI spectrum (Figure~\ref{fig:profiles}).  Each coronal line contains two velocity components: ${\rm [Fe\,X]}\,\lambda 6375$, ${\rm [Fe\,XI]}\,\lambda 7894$, and ${\rm [S\,XII]}\,\lambda 7612$ have velocity separations of $215\pm8$, $240\pm2$, and $230\pm6$ km s$^{-1}$, respectively. These velocities are roughly consistent within uncertainties ($\lesssim 3\sigma$ variation). The narrow component widths are consistent with the coronal line-emitting gas residing at ${\sim}0.8$ pc from the SMBH, which is consistent within a factor of a few with the constraints on the position of the MIR emitting dust, as will be described in Section~\ref{sec:IR}.

Coronal lines in Seyferts are typically blueshifted \citep[][]{Gelbord2009}. The blueshift is thought to indicate the ubiquitous presence of radiatively driven outflows from the AGN torus \citep[][]{Gelbord2009}. In contrast, we observe both a red- and blueshifted component with roughly equal flux. Moreover, the linewidths of coronal lines in Seyferts are often broader than the [O\,III] linewidth, whereas we observe narrower coronal line emission. No other ECLE has a published optical spectrum with sufficiently high resolution to decompose the line profiles, although the coronal lines sometimes appear non-Gaussian in the available, low-resolution spectra \citep[][]{Wang2012}. By eye, the published line profiles seem inconsistent with two, equal-flux peaks.

The coronal line gas could be entrained in and accelerated by the synchrotron emitting outflow (see Section~\ref{sec:radio}), but the line widths are too narrow and the velocity difference between the components too small to favor this scenario. Alternatively, we may be observing rotating gas clouds at a radius ${\sim}0.3$ pc, or an obscured, gaseous disk. The coronal line emitting clouds could also be moving in a radiation-driven outflow, as is thought to occur in Seyferts \citep[][]{Gelbord2009}. We tentatively favor the final scenario although, as we discussed above, the observed line profiles are different from those in typical Seyferts given that the radiatively driven outflow model has observational support in coronal-line-emitting Seyferts and the different profiles could result from a different geometry. Future observations of the line profile evolution and more detailed modelling, such as was done in \cite{Mullaney2009} using \texttt{CLOUDY}, would constrain this scenario.

Next, we constrain the physical properties of the emitting region. We assume the emission is dominated by photoionized gas. This is a reasonable assumption because shocks only strongly contribute to coronal line emission for shock velocities $\gtrsim 300$ km s$^{-1}$, which is much larger than the coronal line widths \citep[][]{Viegas1989}. Photoionized gas is expected to be at a temperature ${\sim}10^5$ K \citep[][]{Korista1989}, so we adopt this as our fiducial value.

First, we roughly estimate the emission measure of the gas following \cite{Wang2011}. Because the gas is photoionized, it must be optically thin to bound-free absorption of soft X-rays of energy ${\sim}1$ keV. We ignore collisional de-excitation to simplify the calculations. For a uniform emitting region of volume $V$ with an ion (electron) density $n_{i(e)}$, the emission measure is given by ${\rm EM} = n_e n_i V$. For a given ion $i$, The emission measure is related to the observed line luminosity $L_i$: $EM = L_i / C_i(T)$. Here, $C_i(T)$ is the collisional strength for the relevant ion at the gas temperature $T\sim10^5$ K. We retrieve the collision strengths from the CHIANTI archive \citep[][]{Dere1997, DelZanna2021}. We find that the emission measures for each strong coronal line are similar, with ${\rm EM}_{\rm CL} \sim 10^{58-59}$ cm$^{-3}$. Assuming the gas has solar abundances, the sulfur and iron abundances are both $n/n_H \sim 10^{-5}$ \citep{Draine2011_book}. We assume both sulfur and iron are dominantly in the observed highly ionized states. Then, we can write $n_e n_H V \sim n_H^2 V \sim 10^{64}$ cm$^{-3}$ and
\begin{equation}
    V = \frac{4}{3}\pi R^3 = 10^{46}\,{\rm cm}^{3}\bigg(\frac{n_H}{10^9\,{\rm cm}^{-3}}\bigg)^{-2},
\end{equation}
\begin{equation}
    R = 1.3 \times 10^{15}\,{\rm cm}\bigg(\frac{n_H}{10^9\,{\rm cm}^{-3}}\bigg)^{-2/3},
\end{equation}
\begin{equation}
    M = m_H n_H V = 8.4 \times 10^{-3}\,M_\odot\,\bigg(\frac{n_H}{10^9\,{\rm cm}^{-3}}\bigg)^{-1}.
\end{equation}

We adopt a distance ${\sim}0.8$ pc based on the coronal line widths. The coronal line-emitting gas may be at a different distance if it is outflowing, but given the low velocity we do not expect the true distance to be changed by more than a factor of a few. With this assumption, the gas must have $R \lesssim 0.8$ pc or $n_H \gtrsim 10^4$ cm$^{-3}$. The detection of [Fe\,X] emission requires $n_H \lesssim 10^9$ cm$^{-3}$, which is the critical density of that line. This density range corresponds to $ 8.4\times10^{-3}\,M_\odot \lesssim M \lesssim 800$. The large mass at the upper bound leads us to favor a higher density than ${\sim}10^4$ cm$^{-3}$. The gas column density is $10^{22} \lesssim N_H/{\rm cm}^{-2} \lesssim 10^{24}$.

For column densities above a few times $10^{23}$ cm$^{-2}$ the gas is optically thick to X-rays. We require optically thin gas. If the gas is clumpy or in a thin shell, the column density will be scaled by a factor of $\xi^{2/3}$, where $\xi = \Delta R/R$ is the relative thickness of the shell or clumps. Likewise, the radius will scale by a factor of $\xi^{-1/3}$. If we adopt a column density ${\sim}10^{22}$ cm$^{-2}$, we find $\xi \sim 10^{-(3-5)}$. Thus, either the coronal line-emitting gas has a very low density but fills a large volume, which is unlikely given the distinctly double peaked, narrow line profiles, or it is dense with a low covering factor. We favor the latter scenario.

Finally, we can constrain the soft X-ray flux required to power the coronal emission. In coronal line-emitting Seyferts, the coronal line luminosity is correlated with the flux in the soft X-ray photoionizing continuum \citep[][]{Gelbord2009}. If we assume that ECLEs lie on this correlation, we can extrapolate to the required soft X-ray flux to power the observed coronal line emission. Given the Seyfert relation $\log f_{\rm Fe\,X} / f_{\rm X} = -3.43 \pm 0.55$ \citep{Gelbord2009}, where $f_{\rm Fe\,X}$ is the flux in the [Fe\,X] line and $f_{\rm X}$ is the X-ray flux, we require a soft X-ray luminosity ${\sim}3\times 10^{42}$ erg s$^{-1}$. We will discuss the origins of this flare in more detail in Section~\ref{sec:discussion}.

In summary, we have detected strong, double-peaked coronal line emission (comparable to the [O\,III] emission). The emission likely comes from clumped gas accelerated by a radiatively driven wind or orbiting the SMBH at ${\sim}0.3$ pc. The coronal lines require an X-ray source with luminosity ${\sim}3\times 10^{42}$ erg s$^{-1}$, with significant uncertainty.

\subsection{Broad Balmer emission} \label{sec:broad}

\begin{figure*}
    \centering
    \includegraphics[width=\textwidth]{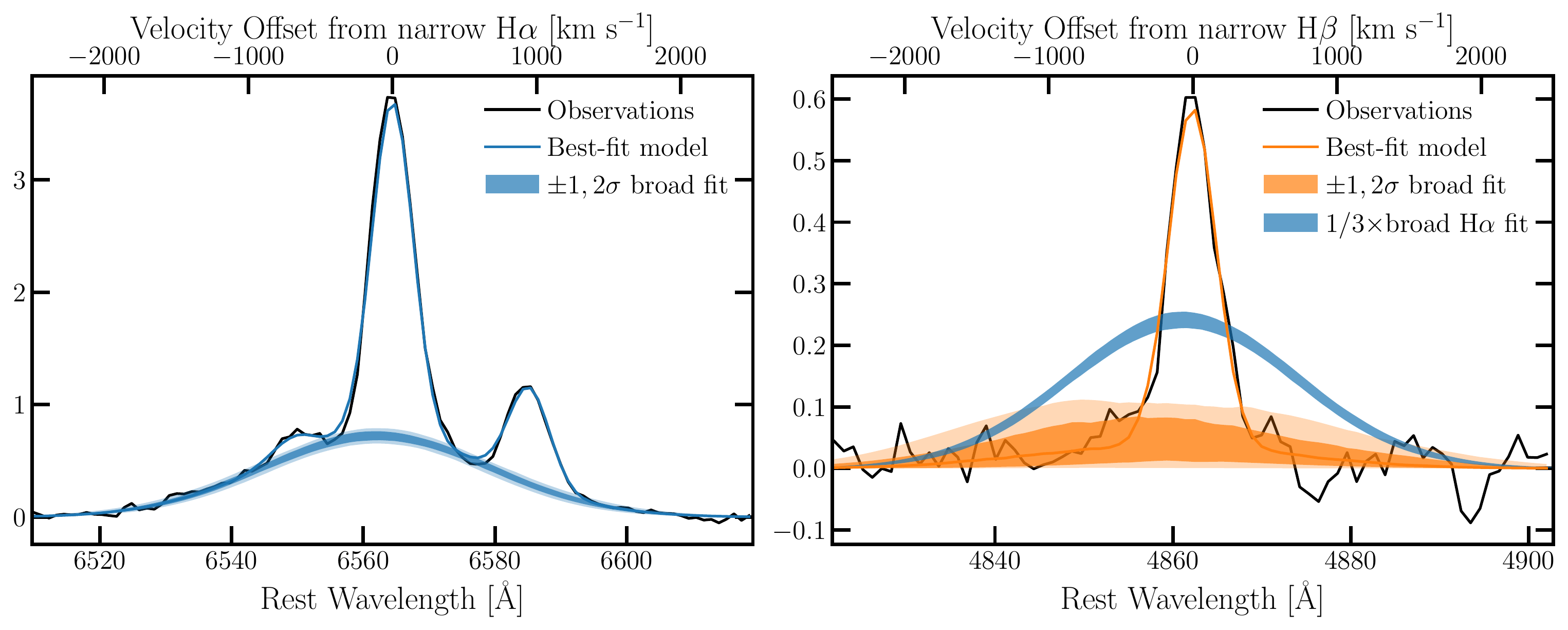}
    \caption{Balmer line profiles from the second epoch of LRIS observations of SDSS J1548. The {\it left} panel shows the H$\alpha$ and N\,II line profiles. The faint lines show the observations and the solid lines show Gaussian fits with uncertainties. A strong, broad H$\alpha$ component is clearly present. The {\it right} panel shows the H$\beta$ profile and fit. We overplot the broad H$\alpha$ fit scaled down by the expected H$\alpha$/H$\beta$ ratio $\sim 3$ \citep[][]{Dong2008}. The H$\beta$ profile is inconsistent with including such a strong, broad component, suggesting that the broad emission must be heavily extincted. }
    \label{fig:broad}
\end{figure*}

VT J1548 is associated with strong, broad H$\alpha$ emission with luminosity ${\sim} 4 \times 10^{39}$ erg s$^{-1}$ and width ${\sim}1900$ km s$^{-1}$ (Figure~\ref{fig:broad}). The detection of late time broad H$\alpha$ from an optical TDE is uncommon, but this luminosity and width are both consistent with upper limits on ${\sim}1000$ day H$\alpha$ emission in optical TDEs (see Figure 7 of \citealp{Brown2017}). The ${\sim}1900$ km s$^{-1}$ width corresponds to a radius $\sim 5\times10^{-3}$ pc $\sim 1000$ AU whether the gas is orbiting the SMBH or driven in an outflow (the correspondence between the expected radius in each case is a coincidence).

The H$\alpha$ luminosity is dimmer than typical AGN emission. The \cite{Greene2005} relationship between SMBH mass and the broad H$\alpha$ luminosity/width predicts $\log M_{\rm BH}/M_\odot \sim 5.3$ from the observed broad H$\alpha$, which is smaller than the SMBH mass predicted by the $M_{\rm BH}-\sigma_*$ relation. The \cite{Greene2005} relation was not calibrated to such low mass BHs and this line is heavily extincted (see next paragraph), so we cannot exclude that the broad emission is consistent with or brighter than that from AGN.

We see no evidence for broad H$\beta$, which is unusual for optical TDEs. In AGN broad line regions (BLRs), the expected value of the H$\alpha$/H$\beta$ ratio is universally ${\sim}3$ \citep[][]{Dong2008} so we expect an H$\beta$ luminosity $\sim 2 \times 10^{39}$ erg s$^{-1}$. Possible modifications to account for collisional excitation can increase the ratio to ${\lesssim}5$, although whether these higher ratios are ever observed is debated \citep[e.g.][]{Ilic2012}. As shown in Figure~\ref{fig:broad}, such a bright line would be detectable. 

Extinction in the galactic nucleus preferentially obscures broad H$\beta$ because it is bluer than H$\alpha$. Extinction is related to the Balmer line ratio as $E(B-V)_{\rm nuc} = 1.97 \log \frac{{\rm H}\alpha/{\rm H}\beta}{3}$. We set an upper limit on the broad H$\beta$ flux by force-fitting a Gaussian profile at the location of H$\beta$ with a FWHM constrained to be within $1\sigma$ of the broad H$\alpha$ FWHM. The $3\sigma$ lower limit on the extinction is $E(B-V)_{\rm nuc}>0.7$. Comparing to measurements of column density and dust extinction in Seyferts BLRs \citep{Schnorr2016}, we find that this corresponds to an absorbing column density $\log N_H/{\rm cm}^{-2} \gtrsim 21.5$. 

This extinction is similar to that observed in Seyfert 1.9 galaxies \citep{Schnorr2016}. Seyfert 1.9s are an inhomogeneous class \citep[see][and references therein]{Hernandez2017}. Some fraction  likely have a large torus that extincts the broad H$\beta$. Galactic-scale extinction can also play a role in these Seyferts, as well as an abnormal nuclear continuum. As discussed in Section~\ref{sec:discussion}, we favor a torus as the cause of the high extinction in VT J1548. We cannot exclude the latter two possibilities. For example, a dust lane covering the nucleus could obscure the BLR while remaining consistent with the observed narrow Balmer decrement ${\sim}3.3$ if the narrow H$\alpha$ comes from a very extended region.

In summary, we detect broad H$\alpha$ but no broad H$\beta$, suggesting we are observing high velocity gas near the SMBH through a screen of obscuring material. 

\section{Analysis of transient broadband features} \label{sec:transientbroad}

\begin{figure*}
    \centering
    \includegraphics[width=\textwidth]{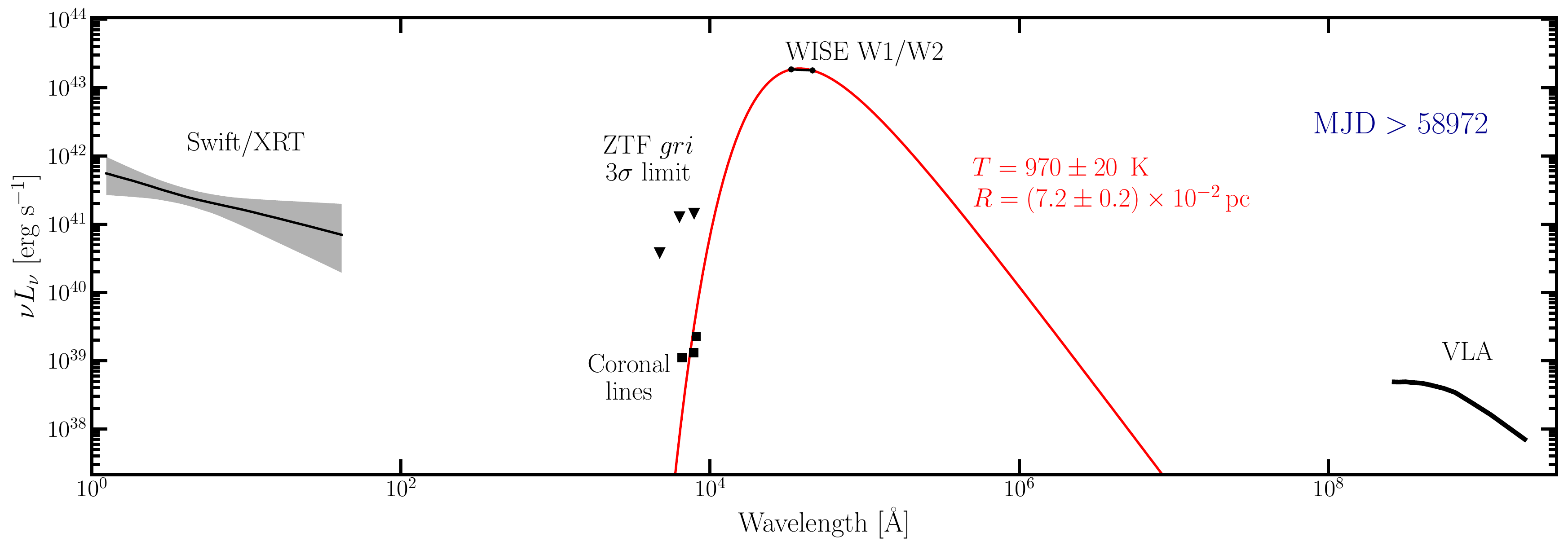}
    \caption{Late time multiwavelength SED of VT J1548. We show the most recent WISE and ZTF obsrvations, along with the VLA SED and the unabsorbed Swift/XRT best-fit spectrum. We also include the coronal line emission and the best-fit blackbody to the WISE emission. Note that the reported blackbody parameter uncertainties are only due to errors in the NEOWISE flux measurement; differential internal extinction between the WISE W1 and W2 bands could increase the temperature by ${\sim}250$ K for the extreme case $E(B{-}V)=3$. }
    \label{fig:sed}
\end{figure*}

VT J1548 was associated with flares in the infrared (Section~\ref{sec:IR}), radio (Section~\ref{sec:radio}), and X-ray (Section~\ref{sec:xray}). The light curve for each flare is shown in Figure~\ref{fig:summary}. VT J1548 was not detected in the optical, and we postpone discussion of the non-detection to Section~\ref{sec:discussion}.

\subsection{Infrared flare} \label{sec:IR}

VT J1548 is associated with a bright ($\Delta m \sim 2$), long lasting (${\gtrsim}1000$ day) flare in the WISE MIR bands. This flare was $>5\times$ brighter than the quiescent state variability. Recent work on IR flares in galactic nuclei has largely argued that the flares can be modeled as ``dust echoes'' \citep{Lu2016}. Dust echoes occur when X-ray/UV photons are absorbed by circumnuclear dust and reprocessed into IR emission. 

Dust echo emission can be fit using detailed models including the dust geometry and emission properties, but they typically agree closely with a blackbody fit \citep[e.g.][]{Kool2020}. We fit a blackbody curve to the WISE data points at each epoch. Figure~\ref{fig:sed} shows the WISE SED and blackbody fit in the final epoch. We only report uncertainties due to the flux errors reported by NEOWISE. We emphasize that these uncertainties do not account for internal extinction: while extinction is small in the WISE bands, differential extinction between the W1 and W2 bands could increase the measured blackbody temperature by as much as ${\sim}250$ K for an extreme $E(B{-}V)=3$. This shift is sufficiently small that it does not change our conclusions significantly but should be noted.

\begin{figure*}
    \centering
    \includegraphics[width=\textwidth]{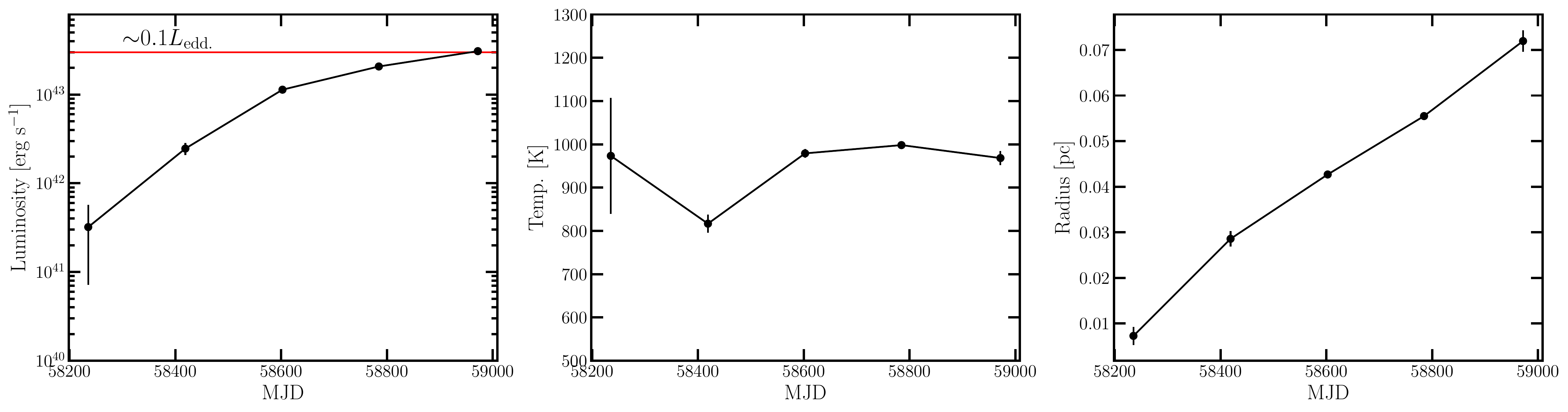}
    \caption{The best-fit blackbody parameters for the mid-IR transient associated with VT J1548. The luminosity ({\it left}), temperature ({\it middle}), and radius ({\it right}) evolution are shown with $1\sigma$ uncertainties. The reported uncertainties are due to errors in the NEOWISE flux measurement. Differential internal extinction between the WISE W1 and W2 bands could increase the temperature by ${\sim}250$ K for the extreme case $E(B{-}V)=3$.}
    \label{fig:TR_evol}
\end{figure*}

The emission plateaus at a near constant temperature ${\sim}1000$ K (Figure~\ref{fig:TR_evol}). The blackbody radius grows from $0.7\times10^{-2}$ pc to $7\times10^{-2}$ pc (although note that this radius does not correspond to the size of the emitting region but instead encodes information about the dust geometry and properties, see discussion in the rest of this section). The dust luminosity has risen to ${\sim}3\times 10^{43}$ erg s$^{-1} \sim 0.1 L_{\rm edd.}$ and has yet to fade.

Integrating the blackbody flux, we find a lower limit on the total emitted energy ${\sim} 5\times10^{50}$ erg. If we assume that this energy is provided by accretion with an efficiency $\eta \sim 0.1$, the accreted mass must be ${\gtrsim}10^{-3}\,M_\odot$. This is consistent the energy emitted during the first few hundred days of typical TDEs, although a factor of $10-100$ more energy may be emitted on much longer timescales (${\gtrsim}5$ years) \citep[see][for a review]{vanVelzen2019}.

A simple explanation of the rising light-curve and nearly constant temperature is a light-travel delay due to dust on different sides of the SMBH. This means that the dust is located at a distance ${\sim}1000\,{\rm day}\times c / 2 \sim 0.4$ pc from the source. We can determine the bolometric luminosity required to produce the IR radiation using the equilibrium between heating and radiative cooling:
\begin{equation}
    e^{-\tau} \frac{L_{\rm bol}}{4\pi R^2} \pi a^2 Q_{\rm abs} = \langle Q_{\rm abs} \rangle_{\rm P} 4 \pi a^2 \sigma T^4.
\end{equation}
$\tau$ is the optical depth for absorption of the heating photons at radii $<R$. $L_{\rm bol}$ is the bolometric luminosity of the flare. $R$ is the emitting radius and $T$ is the emitting temperature. $a_\mu$ is the grain size in units of microns. $Q_{\rm abs} \sim 1$ is the absorption efficiency for the incident photons \citep[][]{Draine2011}, while $\langle Q_{\rm abs} \rangle_{\rm P} \sim a_\mu (T/1000 K)$ is the Planck-averaged absorption efficiency appropriate for $a_\mu \lesssim 1$ and $500 \lesssim T/{\rm K} \lesssim 1500$ \citep[][]{Draine1984}. Assuming a negligible optical depth $\tau$, we find the bolometric luminosity of the flare is $L_{\rm bol} \sim 10^{44}$ erg s$^{-1}\sim L_{\rm edd.}$ assuming a grain size of $0.1$ micron \citep[][]{Draine1984}. The flare was due to a near- or super-Eddington episode of accretion.

Alternatively, we can estimate the bolometric luminosity required to heat the dust from the total emitted energy and rise time. The rise time of the IR emission sets an upper bound on the length of the flare that heated the dust. Given that the luminosity seemed to near a plateau or peak at MJD ${\sim}59000$ (Figure~\ref{fig:TR_evol}), the total length of the ionizing flare is probably $\lesssim 1000$ days. The total emitted energy is ${\sim}5\times10^{50}$ erg s$^{-1}$. If we assume a dust covering factor of ${\sim}1\%$, which is typical of optically selected TDEs \citep[][]{Jiang2021_IRTDE, vanVelzen2016} we find $L_{\rm bol} \gtrsim 6 \times 10^{44}$ erg s$^{-1}$. If we assume a covering factor ${\sim}10\%$, which is consistent with an AGN torus \citep[][]{Ricci2017}, $L_{\rm bol} \gtrsim 6 \times 10^{43}$ erg s$^{-1}$. We favor a higher covering factor (${\gtrsim}10\%$) given the high extinction of the broad line region described in Section~\ref{sec:broad}. Regardless of the covering factor, the UV flare which heated the dust must have been near- or super-Eddington. 

In summary, VT J1548 is associated with a ${\sim}10\%$ Eddington MIR flare that has been ongoing for ${\sim}3$ years. The MIR emission is powered by a near- or super-Eddington nuclear flare.

\subsection{Radio emission} \label{sec:radio}

\begin{deluxetable*}{c|cccc}
\tablecaption{Best-fit Radio SED Parameters \label{tab:radio_sed}}
\tablewidth{0pt}
\tablehead{\colhead{Parameter} & \colhead{SSA} & \colhead{FFA} & \colhead{Inhomogeneous SSA} & \colhead{Multi-comp. SSA} }
\startdata
$K_1$ & $0.46^{+0.02}_{-0.02}$ & $9.65^{+0.3}_{-0.3}$ & $1.85^{+0.07}_{-0.08}$ & $1.68^{+0.28}_{-0.27},3.2^{+0.55}_{-0.71}$ \cr
$K_2$ & $17.5^{+1.2}_{-1.1}$ & $4.1^{+0.2}_{-0.2}$ & $38.5^{+3.9}_{-3.6}$ & $31^{+67}_{-14},278^{+165}_{-59}$ \cr
$\alpha$ & $0.50^{+0.01}_{-0.01}$ & $0.56^{+0.02}_{-0.01}$ & $0.5^{+0.04}_{-0.04}$ & $1.4^{+1.3}_{-0.6},1.06^{+0.18}_{-0.05}$ \cr
$\alpha'$ & $-$ & $-$ & $1.35^{+0.05}_{-0.05}$ & $-$ \cr
$\chi^2/{\rm dof}$ & $496/78$ & $458/78$ & $69/77$ & $62/74$
\enddata
\centering
\tablecomments{All fluxes are assumed to be in mJy and frequencies in GHz. $1\sigma$ uncertainties are reported. The multi-comp. SSA model includes a low frequency pure power law component that is not included in the reported fits, see text for details of the model. }
\end{deluxetable*}

\begin{figure}
    \centering
    \includegraphics[width=0.49\textwidth]{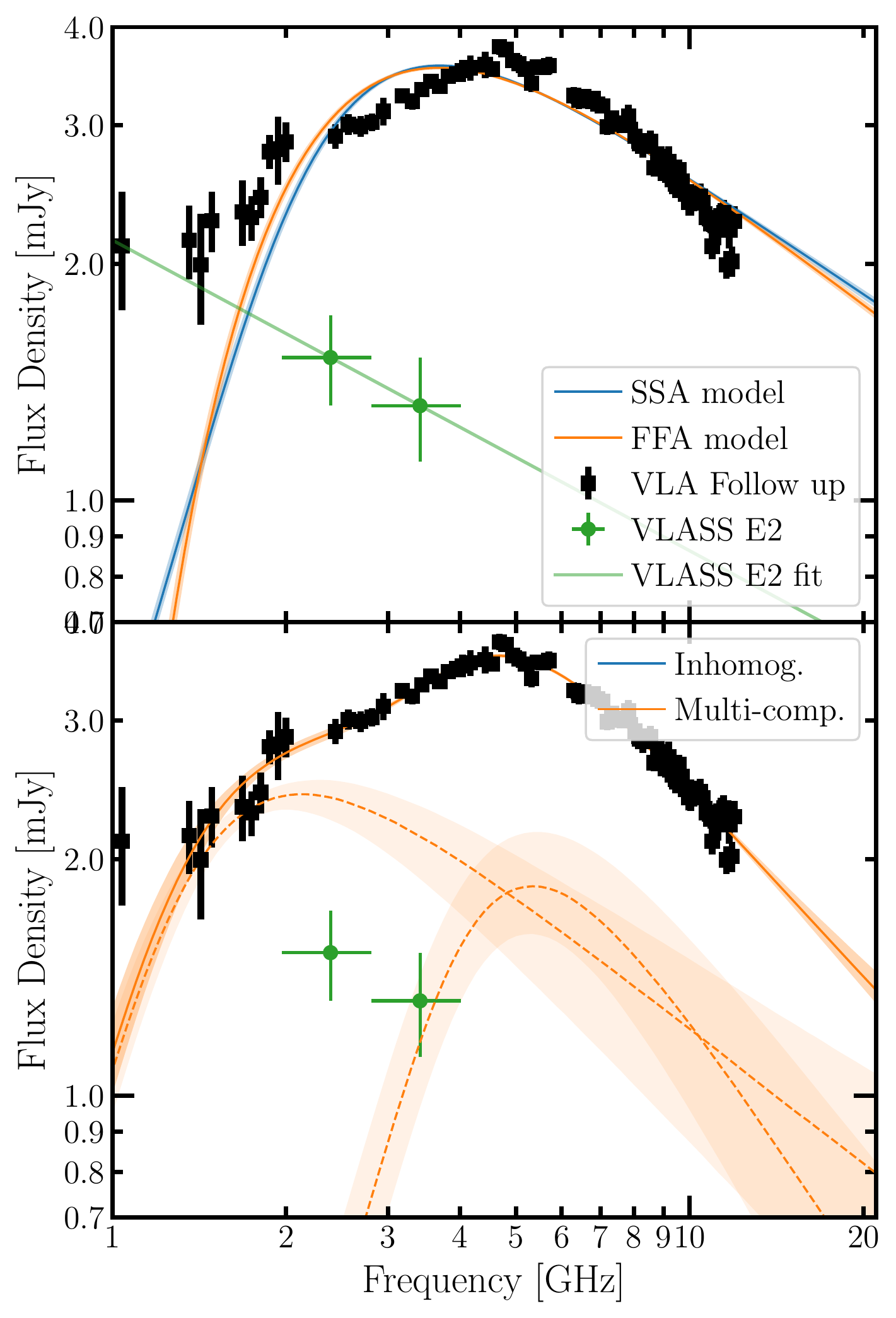}
    \caption{The observed radio SED and best fit models. The {\it top} panel show the VLA follow-up observations in black, the VLASS E2 observations in green, and the best fit self absorbed synchrotron and free-free absorbed models, with $1\sigma$ error bars. In both cases, the models provide extremely poor fits. We also show an extrapolation of a power law fit to the VLASS E2 points in green. The {\it bottom} panel shows the non-standard synchrotron model fits. The blue band shows the best-fit inhomogeneous model. The top-most orange band shows the best-fit multi-component synchrotron model. Each component is shown as an orange band in the lower part of the panel. The non-standard models both provide substantially better fits.}
    \label{fig:radio_sed}
\end{figure}

\begin{deluxetable}{c|cccc}
\tablecaption{Results of the synchrotron equipartition analysis \label{tab:radio_phys}}
\tablewidth{0pt}
\tablehead{\colhead{Parameter} & \colhead{Low freq. Comp.} & \colhead{High freq. Comp.} }
\startdata
$p$ & $2.28^{+0.37}_{-0.20}$ & $3.19^{+0.47}_{-0.35}$ \cr
$\log R_p/{\rm cm}$ & $17.00^{+0.06}_{-0.03}$ & $16.68^{+0.04}_{-0.03}$ \cr
$\log U_p/{\rm erg}$ & $48.6^{+0.3}_{-0.08}$ & $48.77^{+0.4}_{-0.2}$ \cr
$B_p$ [G] & $0.13^{+0.025}_{-0.009}$ & $0.469^{+0.13}_{-0.08}$ \cr
$\beta$ & $0.065^{+0.09}_{-0.005}$ & $0.03^{+0.003}_{-0.002}$  \cr
$\log n_e/{\rm cm}^{-3}$ & $2.18^{+0.17}_{-0.08}$ & $3.19^{+0.47}_{-0.35}$ \cr
\enddata
\centering
\tablecomments{All parameters are derived assuming assuming equipartition with $\epsilon_e = \epsilon_B = 0.1$. We assume that both the low and high frequency components correspond to outflows that launched ${\sim}700$ days after the beginning of the IR flare.}
\end{deluxetable}

In this section, we discuss the transient radio emission. First, we consider the rapid light curve evolution. Then, we model the broadband SED.

The radio light curve is shown in Figure~\ref{fig:summary}. If the radio emission turned on when the IR emission turned on, the fast rise between the VLASS E2 and VLA follow-up observations requires $F_\nu ({\rm 3\,GHz}) \propto \Delta t^{4.5}$. The fastest expected optically thick flux density rise is $F_\nu ({\rm 3\,GHz}) \propto \Delta t^{3}$ for an on-axis, relativistic jet, which is likely an oversimplification \citep[see discussion in][]{Horesh2021}. The observed radio emission rises as $\Delta t^3$ if it turned on ${\sim}400$ days after the IR flare (MJD 58530). The emission is best modeled as sub-relativistic (see discussion at the end of this section), so the light curve should rise more slowly than $\Delta t^{2.5}$, which corresponds to an outflow launch date ${\sim}500$ days after the initial IR flare (MJD 58635). These rise times all assume a constant circumnuclear density profile, which is likely incorrect. The launch date need not be delayed if the outflow evolved for ${\sim}700$ days before colliding with a dense shell of material.

The radio SED provides insight into the unusual light curve evolution. The observed SED, shown in Figure 1, has evolved significantly between the VLASS E2 observations (green points) and the VLA follow up (black points). The uncertainty on the in-band slope from the VLASS E2 observations is too large to make any conclusive claims, but the $2-3$ GHz slope has stayed roughly flat.

Radio emission from a TDE may result from a relativistic or sub-relativistic outflow interacting with the circumnuclear material (CNM) and producing a synchrotron-emitting shockwave. We assume the emission is produced by a population of electrons with a power law energy distribution:
\begin{equation} \label{eq:el_en}
    \frac{dN(\gamma)}{d\gamma} \propto \gamma^{-p}, \gamma \ge \gamma_{\rm m}.
\end{equation}
The index $p$ depends on the acceleration mechanism, with typical mechanisms producing $2\lesssim p \lesssim 3$. The minimum electron Lorentz factor, $\gamma_{\rm m}$, is set by $\epsilon_e$, the fraction of the total energy used to accelerate electrons. Equipartition is commonly assumed: $\epsilon_e = \epsilon_B \sim 0.1$, where $\epsilon_B$ is the fraction of the energy density stored in magnetic fields. The SSA model includes characteristic frequencies: $\nu_m$, $\nu_{sa}$, and $\nu_c$. $\nu_m$ is the synchrotron frequency of the minimum energy electrons. $\nu_{sa}$ is the frequency below which emission is optically thick so synchrotron self-absorption (SSA) is important. $\nu_c$ is the cooling frequency where the electron age is equal to the characteristic cooling time by SSA. We refer the reader to \cite{Ho2019} for a concise and clear description of SSA models and the characteristic frequencies.

Typically, the dominant absorption mechanism in TDE-driven outflows is SSA. Then, the radio flux density can be written \citep{Snellen1999}
\begin{gather} \label{eq:radiosed}
    \frac{F_\nu}{\rm mJy} = K_1 \bigg( \frac{\nu}{1\,{\rm GHz}} \bigg)^{2.5}  (1-e^{-\tau_{\rm SA}}). \\
    \tau_{\rm SSA} = K_2 \bigg(\frac{\nu}{1\,{\rm GHz}}\bigg)^{-(\alpha+2.5)}.
\end{gather}
$K_{1,2}$ are normalizations characterizing the SED flux and optical depth, respectively. $\alpha$ is the optically thin slope. $\tau_{\rm SSA}$ is the optical depth to SSA. We are forcing the optically thick slope to be $5/2$, which is expected for optically thick blackbody emission, where the blackbody temperature depends on frequency as $\nu^{1/2}$.

We fit this model to the observations using the \texttt{dynesty} dynamic nested sampler \citep[][]{Speagle2020} with uninformative Heaviside priors. The best-fit SED is shown in the top panel of Figure~\ref{fig:radio_sed}, and the best-fit parameters are summarized in Table~\ref{tab:radio_sed}. The observed optically thick slope is shallower than the canonical $5/2$. Variations on this standard SSA model can predict slopes as shallow as $2$ \citep{Granot2002}, which is still inconsistent with our observations.

One possible modification of this model is strong free-free absorption (FFA) rather than SSA. The SED for an FFA dominated model is \citep[][]{Chevalier1998}:
\begin{gather} \label{eq:radiosedFFA}
    \frac{F_\nu}{\rm mJy} = K_1 \bigg( \frac{\nu}{1\,{\rm GHz}} \bigg)^{-\alpha}  e^{-\tau_{\rm FFA}}. \\
    \tau_{\rm FFA} = K_2 \bigg(\frac{\nu}{1\,{\rm GHz}}\bigg)^{-2.1}.
\end{gather}
We fit this FFA model to the observations using the same techniques as for the SSA model. The best fit parameters are tabulated in Table~\ref{tab:radio_sed} and the model is shown in Figure~\ref{fig:radio_sed}. The fit is poor with $\chi^2/dof = 458/78$.

We may not be in the canonical regime with $\nu_{sa} < \nu_m < \nu_c$ for which the above parameterizations apply. As we discuss later in this section, the magnetic fields consistent with our SED are ${\sim}0.5$ G. Assuming a ${\gtrsim}500$ day age of the emission, the corresponding cooling frequency is higher than our highest frequency observation, whereas the other two characteristic frequencies are much smaller. Instead, we must consider non-standard emission models. First, we use a model that allows for inhomogeneities in the emitting region. Then, we consider the sum of multiple, independent SSA models. 

We model an inhomogenous emitting region following \cite{Bjornsson2013, Bjornsson2017, Chandra2019}. The probability of observing a given magnetic field is $P(B) \propto B^{-a},\,B_0 < B < B_1$. When the frequency is below the characteristic synchrotron frequency at $B_0$, the SED will have the standard optically thick slope of $5/2$. The slope for frequencies above the synchrotron frequency for $B_1$ is interpreted as the optically thin slope in the standard SSA model. In between, the SED slope is $\alpha' = (3p +5\delta' - a(p+4))/(p+2(1+\delta'))$, where $0 \le \delta' \le 1$ characterizes a correlation between the electron distribution and the magnetic field strength distribution, and all other variables are as defined earlier. We assume the optically thick region with slope $5/2$ is at frequencies lower than our observations, and adopt the model:
\begin{gather} \label{eq:ihradiosed}
    \frac{F_\nu}{\rm mJy} = K_1 \bigg( \frac{\nu}{1\,{\rm GHz}} \bigg)^{\alpha'}  (1-e^{-\tau_{\rm SA}}). \\
    \tau_{\rm SSA} = K_2 \bigg(\frac{\nu}{1\,{\rm GHz}}\bigg)^{-[\alpha'+(p-1)/2]}.
\end{gather}
The best fit slopes (Table~\ref{tab:radio_sed}) are $\alpha' = 0.5$ and $\alpha=(p-1)/2=1.35$. The value of $\alpha' = 0.5$ corresponds to $a=1.1, 1.6$ for $\delta'=0,1$ respectively. The high frequency spectral slope corresponds to $p\sim3.7$, which is substantially higher than the typical $p<3$. The large $p$ may be unphysical and suggests the inhomogeneities are more complex than assumed.

We conclude our radio SED modelling by fitting the sum of two independent SSA models. The best-fit parameters for each SSA profile are shown in Table~\ref{tab:radio_sed} and the best fit model is shown in the lower panel of Figure~\ref{fig:radio_sed}. The optically thin slopes correspond to $p \sim 2.3/3.2$ for the low and high frequency components, respectively. Both slopes are consistent with $2<p<3$ within $1\sigma$, so we can use a standard equipartition analysis to map the two SSA components to physical parameters of the outflow.

\cite{Chevalier1998} provides a detailed overview of equipartition analyses. In brief, the outer radius of the shock is given by
\begin{equation}
    R_p = \Bigg[ \frac{6 c_6^{p+5} F_p^{p+6} D^{2p+12}}{(\epsilon_e/\epsilon_B) f (p-2) \pi^{p+5} c_5^{p+6} E_l^{p-2}} \Bigg]^{1/(2p+13)} \Bigg( \frac{ \nu_p}{2 c_1} \Bigg)^{-1},
\end{equation}
where the electron rest mass energy $E_l = 0.51$ MeV, $f$ is the filling factor, and  $c_1 = 6.27 \times 10^{18}$ (cgs). $c_5$ and $c_6$ are both functions of $p$ \citep{Pacholczyk1970}. $\nu_p$ is the peak frequency and $F_p$ is the peak flux density. We have adopted the notation of \cite{Ho2019}. 

Assuming a time $t_p$ since the initial event, the speed of the shock is given by $v \sim R_p/t_p$. As discussed at the beginning of this section, the radio light curve for VT J1548 is inconsistent with the dominant synchrotron components corresponding to outflows that are launched with the IR flare. Hence, we calculate the launch date assuming a $t^{2.5}$ rise, so $t_p\sim 600$ days. A smaller $t_p$ would result in a slightly, but not significantly, higher velocity and lower electron density.

Using the same notation, the magnetic field is given by
\begin{equation}
    B_p = \Bigg[ \frac{36 \pi^3 c_5}{(\epsilon_e/\epsilon_B)^2 f^2 (p-2)^2 c_6^3 E_l^{2(p-2)} F_p D^2} \Bigg]^{\frac{2}{2p+13}} \Bigg( \frac{ \nu_p}{2 c_1} \Bigg).
\end{equation}
Here, $D$ is the distance to the source (137 Mpc for SDSS J1548).

Finally, the equipartition energy, which is a lower bound on the true energy, is
\begin{equation}
    U = \frac{1}{\epsilon_B}\frac{4\pi}{3} f R^3 \Bigg(\frac{B^2}{8 \pi} \Bigg)
\end{equation}

The physical parameters for each component are listed in Table~\ref{tab:radio_phys}. Both components are consistent with an energetic, non-relativistic outflow moving through a dense medium. The lower frequency component, which dominates the fast rising light curve, is faster, at slightly larger radius, and is consistent with a lower density than the higher frequency, subdominant component. 

These observations might suggest that the outflow is colliding with an asymmetric and/or inhomogeneous medium. We will discuss this interpretation in Section~\ref{sec:discussion}. In principle, it should be possible to devise a more realistic synchrotron model that includes a physically motivated parameterization of the circumnuclear medium, but such an analysis is beyond the scope of this paper.

To conclude, VT J1548 shows fast-rising, radio emission that is consistent with an outflow at a radius ${\sim}0.1$ pc that is incident on a inhomogenous medium.

\subsection{X-ray emission} \label{sec:xray}

\begin{figure}
    \centering
    \includegraphics[width=0.49\textwidth]{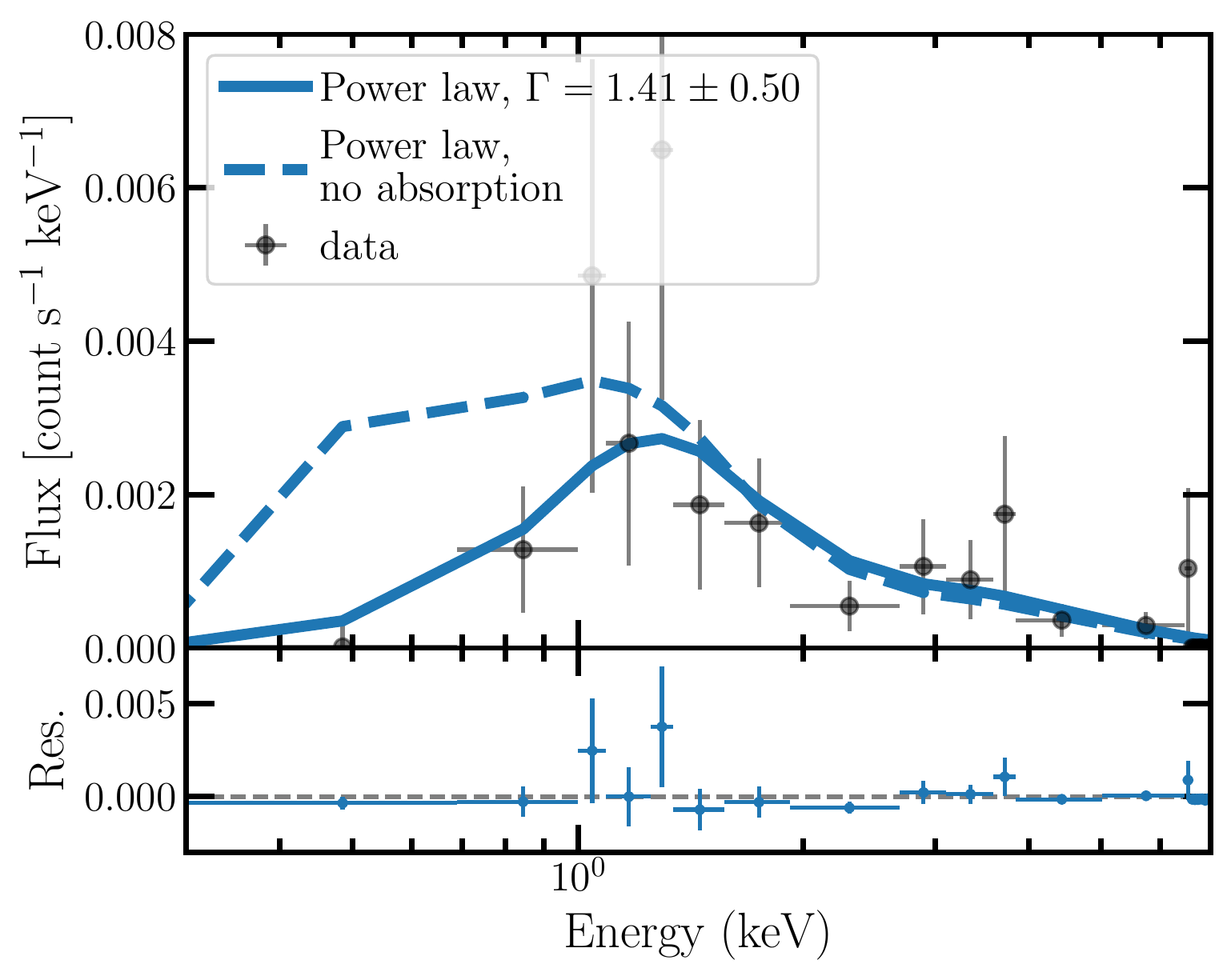}
    \caption{The Swift/XRT X-ray spectrum for VT J1548 (black points) and the best fit power law model (solid blue). The dashed blue line shows the (binned) best-fit model with no absorption.}
    \label{fig:xray_spec}
\end{figure}

\begin{figure*}
    \centering
    \includegraphics[width=\textwidth]{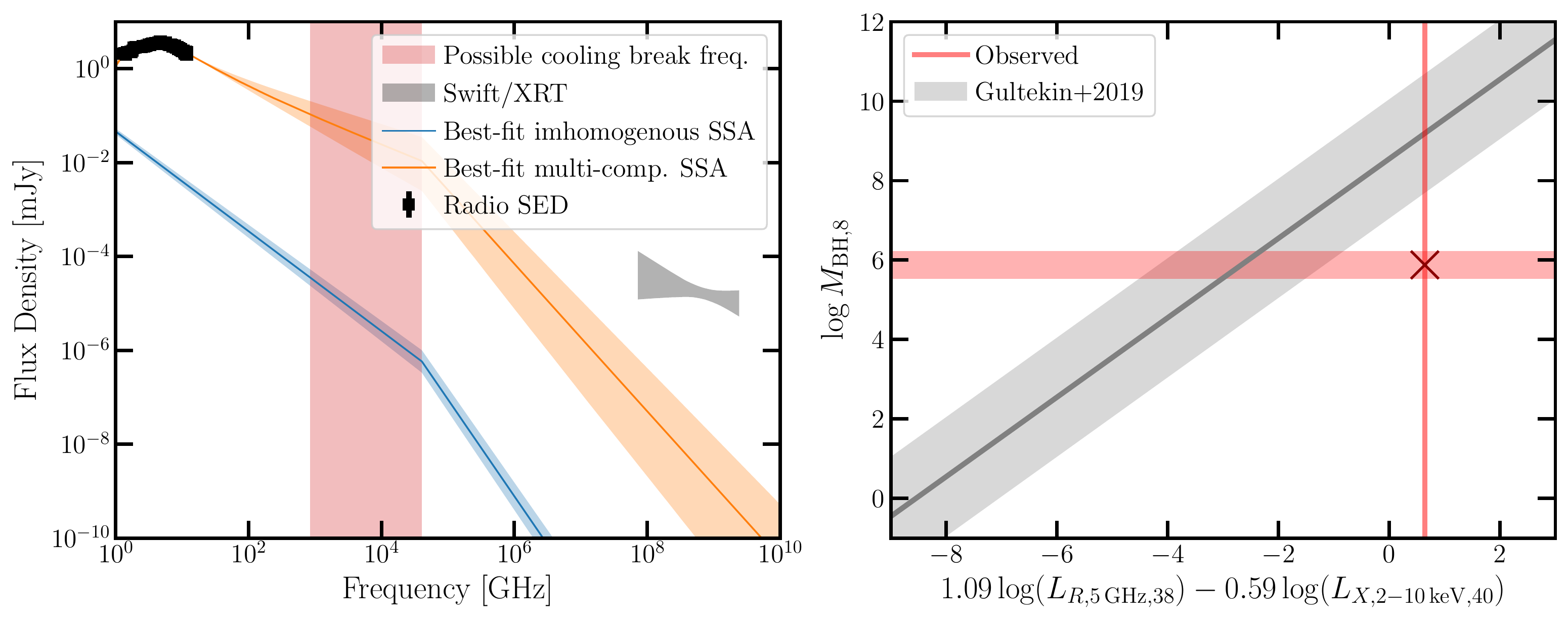}
    \caption{({\it left}) The radio SED compared with the observed X-ray emission. The inhomogeneous (blue) and multi-component (orange) SSA models both predict an X-ray flux that is many orders of magnitude lower than the observations. We have included a cooling break, and the plausible cooling break frequencies are denoted by the red band. The cooling break would have to be at a frequency orders of magnitude higher than predicted for the X-ray synchrotron to be observable. We conclude that it is unlikely that the synchrotron tail contributes the the X-ray emission. ({\it right)} The fundamental plane for black hole accretion from \cite{Gultekin2019}, with observations of SDSS J1548 overplotted. This source is inconsistent with accretion-related emission. }
    \label{fig:xray_radio}
\end{figure*}

Finally, we consider the X-ray emission associated with VT J1548. First, we discuss the X-ray spectrum and luminosity. Then, we consider the source of the X-ray emission.

We model the X-ray spectrum using \texttt{xspec} with \cite{Cash1979} statistics and the \cite{Wilms2000} abundances. We adopt an absorbed, redshifted power law model: \texttt{cflux*TBabs*zTBabs*powerlaw}. We fix the redshift and column density to the known values and allow for extra absorption (\texttt{zTBabs}). The spectrum with the fit overplotted is shown in Figure~\ref{fig:xray_spec}. The best-fit column density is poorly constrained at $n_{\rm H} = (3.7 \pm 5)\times10^{21}$ cm$^{-2}$. The best-fit photon index is $1.41 \pm 0.5$. These parameters are most consistent with AGN observations and late time TDE observations \citep[][]{Auchettl2017, Jonker2020}. They are mildly inconsistent ($3\sigma$) with early time TDE observations \citep[][]{Auchettl2017, Jonker2020, Sazonov2021}.

We used the best-fit power law parameters to convert the PSF, dead time, and vignette-corrected source intensities in the Swift/XRT light curve to a flux using the \texttt{WebPIMMS} tool\footnote{\url{https://heasarc.gsfc.nasa.gov/cgi-bin/Tools/w3pimms/w3pimms.pl}}. As shown in the middle panel of Figure~\ref{fig:summary}, the X-ray emission from SDSS J1548 is bright, with an average Swift/XRT flux $f_{\nu}({\rm 0.2-10\,keV}) = (1.98 \pm 0.33) \times^{-5}$ mJy $=(4.64 \pm 0.78) \times 10^{-13}$ erg cm$^{-2}$ s$^{-1}=(9.9 \pm 1.7)\times10^{41}$ erg s$^{-1}$. This is ${\sim}1\% L_{\rm Edd.}$, and is bright compared to most late-time (${\gtrsim}5$ yr) TDE X-ray detections but probably consistent with ${\sim}1000$ day TDE observations provided that there is on-going accretion years to decades after the event \citep[][]{Jonker2020}. Assuming the X-ray flare has lasted for the same duration as the WISE flare, the total energy output is ${\sim}10^{50}$ erg. 

The observed luminosity is comparable to that required by the coronal lines, so it likely powers the high ionization emission. The X-rays are highly absorbed if we are observing the X-rays through the same material that is obscuring the broad Balmer emission. Assuming a soft spectral slope $\Gamma \sim 3$ and an intrinsic column density $n_H \sim 3\times10^{21}$ cm$^{-2}$, which are consistent with the constraints from the X-ray spectrum and the broad Balmer emission, the intrinsic X-ray luminosity would be ${\sim}$an order of magnitude higher than observed. This emission could originate near the SMBH from, e.g., an AGN-like accretion disk and corona, or the base of a jet.

X-rays can also be emitted by the tail of the radio synchrotron emission. Given the likely presence of a cooling break between the X-ray and radio frequencies, the synchrotron tail underpredicts the observed X-ray emission by orders of magnitude (left panel of Figure~\ref{fig:xray_radio}). Synchrotron emission is not a significant contributor to the X-ray luminosity.

The X-ray flare could be related to normal AGN variability, in which case VT J1548 should lie on the fundamental plane for black holes. In the right panel of Figure~\ref{fig:radio_sed}, we show the fundamental plane from \cite{Gultekin2019} with our observations overplotted. This source is inconsistent with accretion-related emission. Hence, it is unlikely the result of normal (non-extreme) AGN variability.

Inverse Compton scattering of radiation by electrons in the outflow can produce X-rays. In general, the ratio of synchrotron to inverse Compton power is given by
\begin{equation}
    \frac{P_{\rm synch}}{P_{\rm compt}} = \frac{U_B}{U_{\rm ph}},
\end{equation}
where $U_{\rm ph}$ is the photon energy density and $U_B$ is the magnetic field energy density. The magnetic field in the outflow is ${\sim}0.5$ G, so the magnetic energy density is $B^2/(8\pi) \sim 0.01$ erg cm$^{-3}$. The IR luminosity is ${\sim}10^{43}$ erg s$^{-1}$ and is emitted from a radius ${\sim}0.4$ pc. Then, we can set a lower limit on the photon energy density of ${\sim}10^{43}\,{\rm erg\,s}^{-1} \times (0.4\,{\rm pc})/c \times (4/3 \pi (0.4\,{\rm pc})^3)^{-1} \sim 10^{-4}\,{\rm erg\,cm}^{-3}$. Thus, we have $P_{\rm synch}/P_{\rm compt} \sim 100$. The predicted X-ray luminosity from inverse Compton scattering in the outflow is thus ${\sim}10^{36}$ erg s$^{-1}$. This is ${\sim}6$ orders of magnitude lower than observed.

Alternatively, thermal bremsstrahlung in the synchrotron-emitting outflow can produce X-rays. From \cite{Rybicki}, the thermal free-free emissivity (in cgs units) is given by
\begin{equation}
    \epsilon^{ff} = 1.4\times10^{-27} T^{1/2}n_e^2 Z^2 \bar{g}_B(T).
\end{equation}
Here, $T$ is the electron temperature in Kelvin, $\bar{g}_B$ is the velocity and frequency averaged Gaunt factor, which is of order 1. $n_e$ is the electron density in cm$^{-3}$, and we have assumed the electron and ion density are similar. $n_e \sim 10^{3}$ cm$^{-3}$ from our synchrotron model. Assuming a volume of radius ${\sim}0.1$ pc and adopting $n_e\sim10^3$ cm$^{-3}$, the free-free luminosity is $L_{ff} \sim 10^{32} T^{1/2} Z^2$ erg s$^{-1}$. The observed luminosity (${\sim}10^{41}$ erg s$^{-1}$) requires $T\sim10^{18}$ K, which is unrealistically high. The X-ray emission could come from clumps that are much higher density than average but have a low covering factor. For example, the X-rays could be entirely produced by bremsstrahlung from clumps with a temperature $T\sim10^7$ K, $n\sim10^8$ cm$^{-3}$, and a covering factor $\sim0.2\%$. We cannot rule out this scenario.

Hence, we conclude that the ${\sim}10^{42}$ erg s$^{-1}$ X-ray emission likely originates from the same source that is causing the coronal line emission and IR flare, with a possible contribution from thermal bremsstrahlung. As we discuss in the next section, the exact origin of this emission depends on the event that caused the transient. One explanation that could apply in both a TDE scenario or extreme AGN variability is AGN-like soft X-rays from an accretion disk with a hot electron corona, or emission from the base of a nascent jet. A more detailed measurement of the shape of the X-ray spectrum would tighten the constraints on the origin of the X-ray emission.

\section{Discussion} \label{sec:discussion}

In this section, we consider models that explain the emission from VT J1548. First, we summarize the observations of SDSS J1548/VT J1548. Then, we compare VT J1548 to published transients. We present a qualitative cartoon model describing the geometry of the system. Finally, we discuss the possible events that triggered the onset of VT J1548, and we finish by describing observations that could distinguish between these properties and/or clarify our physical model.

The observational properties of VT J1548 and its host, SDSS J1548, can be summarized as follows:
\begin{itemize}
    \item SDSS J1548 is a bulge dominated S0 galaxy. It has line ratios that are marginally consistent with an AGN-like ionizing source. It hosts a low mass black hole, with $\log M_{\rm BH}/M_\odot = 6.48 \pm 0.33$.
    \item VT J1548 is associated with strong (${\rm [Fe\,X]/[O\,III]}\sim1$), double peaked ($\Delta v \sim 230$ km s$^{-1}$) coronal line emission powered by X-ray emission with a luminosity ${\gtrsim}10^{42}$ erg s$^{-1}$.
    \item VT J1548 coincided with the onset of broad H$\alpha$ emission (FWHM $\sim 1900$ km s$^{-1}$), but no broad H$\beta$ emission, suggesting strong internal extinction with $E(B-V) \gtrsim 0.7$.
    \item The transient emission lines commonly associated with optically-selected TDEs (He\,II, N\,III) are undetected. We do not detect any of the [Fe\,II] lines that are abundant in Seyfert spectra.
    \item VT J1548 is associated with a bright ($\Delta m \sim 2-3$) MIR flare. The flare rose over ${\sim}900$ days and had not begun fading from a luminosity of ${\sim}0.1L_{\rm edd.}$ as of MJD 59000. The flare temperature stayed roughly constant at $1000$ K, and the emission is consistent with dust heated by near- or super-Eddington UV flare.
    \item VT J1548 was undetected in the radio shortly before the beginning of the IR flare, but had turned on within ${\sim}2$ years. The radio emission from VT J1548 is currently consistent with an inhomogeneous SSA model or a two-component SSA model peaking at a frequency of 5 GHz with a flux density 4 mJy, although the best-fit parameters for the two-component model are more consistent with theoretical expectations for synchrotron sources. The best-fit parameters suggest the components are both non-relativistic outflows, one of which is slightly faster with a lower electron density and magnetic field.
    \item The transient X-ray emission ($f_X=(4.7\pm0.8)\times 10^{-13}$ erg cm$^{-2}$ s$^{-1}$, $L_X{\sim}10^{42}$ erg s${-1}$, $\Gamma = 1.41 \pm 0.5$) may be AGN-like (i.e., disk and corona). Bremsstrahlung emission from dense clumps of gas may contribute.
\end{itemize}

\begin{figure*}
    \centering
    \includegraphics[width=0.99\textwidth]{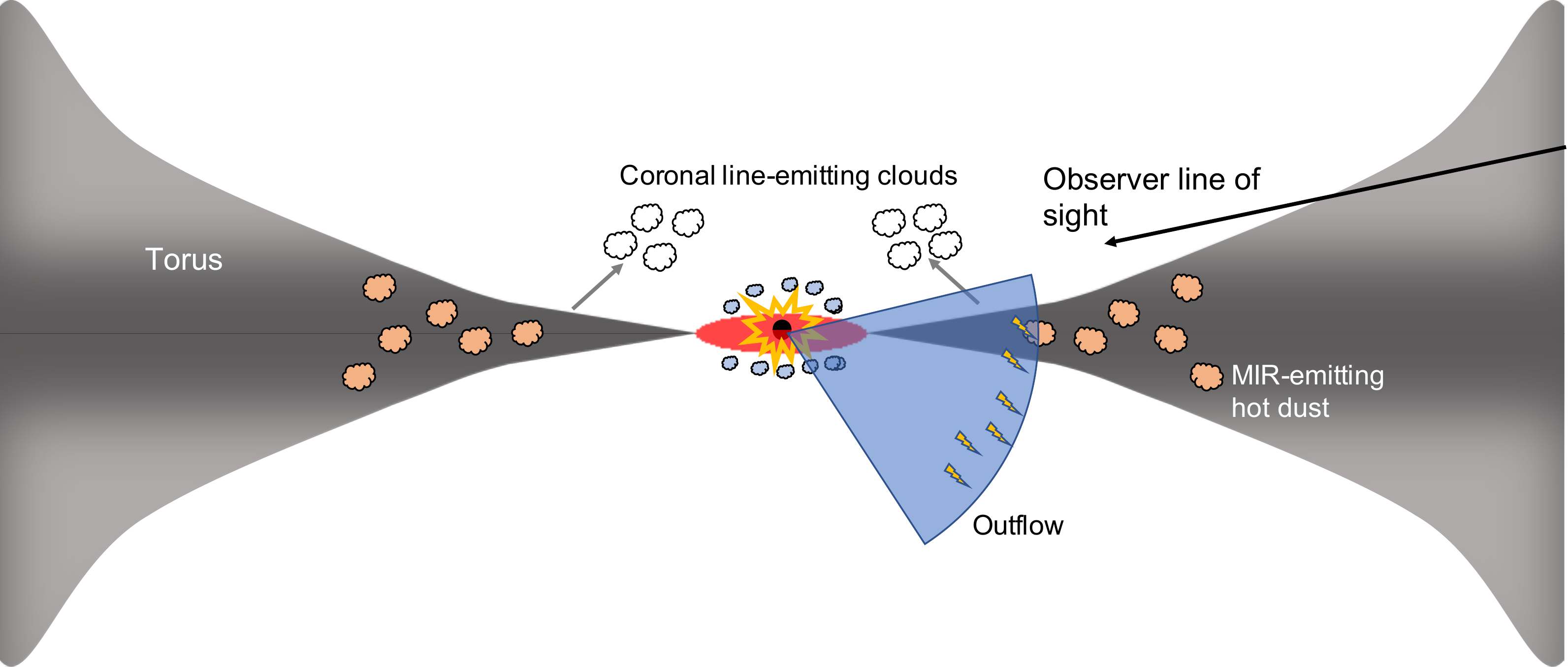}
    \caption{A cartoon showing the approximate geometry of SDSS J1548/VT J1548. Some event triggered a flare at the nucleus of the galaxy, which caused the formation of an accretion disk or was associated with enhanced accretion from a pre-existing accretion disk. The broad H$\alpha$ emission originates from near the accretion disk (${\sim}1000$ AU), and is extincted by the dusty torus along the line of sight to the observer. Any optical emission from the transient event is also heavily extincted. The coronal line-emitting gas, synchrotron-emitting outflow, and MIR emitting dust are all at roughly the same distance (${\sim}0.4$ pc). The coronal line-emitting clouds are embedded in a radiation-driven wind (${\sim}100$ km s$^{-1}$) off of the torus, which causes the double peaked emission. Alternatively, the clouds may be orbiting the SMBH at ${\sim}0.3$ pc to produce the double peaked lines. IR emission comes from the heated dust in the torus. The radio emitting outflow is shown in blue. }
    \label{fig:cartoon}
\end{figure*}

\subsection{Comparison to published transients}

These general features have individually been observed in previous transients, but never together. In this section, we compare VT J1548 to select transients from the literature. We refer the reader to \cite{Zabludoff2021} for a more comprehensive discussion of unusual TDE candidates. In Table~\ref{tab:comp}, we summarize critical properties of VT J1548 and compare them to the ``unusual'' TDE candidates and two unique changing look AGN that we discuss in Section~\ref{sec:agn}. We selected these transients as those that evolve in the optical/IR/X-ray on a timescale slower than the typical TDE ($\gtrsim400$ days) or those that initially evolve on shorter timescales but have late-time ($>400$ day) X-ray detections. In the rest of this section, we highlight some of the unusual transients.

First, tens of extreme coronal line emitters have been observed with coronal line luminosities that are generally a factor of a few higher than that observed from VT J1548 \citep[e.g.][]{Komossa2008, Wang2011, Wang2012, Frederick2019}. High extinction in SDSS J1548 could cause the dim emission. The line profiles from ECLEs have not been studied in detail due to a lack of high resolution follow up, but double peaked profiles are not unprecedented for normal AGN and are generally attributed to a partially obscured rotating disk or an outflow \citep[e.g.][]{Mazzalay2010, Gelbord2009}. It would be unsurprising if high resolution observations of ECLEs uncovered complex line profiles \citep[see][for discussion of possible unusal coronal line profiles in ECLEs]{Wang2012}.

Most ECLEs are inconsistent with past AGN activity whereas SDSS J1548 has line ratios that could be consistent with weak AGN activity \citep[][]{Wang2012}. One exception is AT2019avd \citep[][]{Frederick2020, Malyali2021}, which was selected as an X-ray and optical transient in a galaxy with a low SMBH mass $\log M_{\rm BH}/M_\odot \sim 6.3 \pm 0.3$. Like VT J1548, the host galaxy was consistent with weak or no AGN activity based on archival X-ray non-detections and BPT line ratios. The optical light curve was initially similar to standard, prompt TDE emission (i.e., it evolved over a timescale of ${\sim}100$ days), but it rebrightened significantly $\sim500$ days after the initial peak. Its X-ray emission was very soft ($\Gamma \sim 5$) and the X-ray luminosity ${\sim}600$ days post-optical peak remained at ${\sim}10^{43}$ erg s$^{-1}$, or ${\sim}0.1 L_{\rm edd.}$. It was detected as a WISE flare that turned on after the optical emission. An optical spectrum near the first optical peak showed Fe\,II emission, and another spectrum taken ${\sim}450$ days post-peak showed He\,II and Bowen fluorescence lines. It showed broad transient Balmer emission and a Balmer decrement close to the expected value of $3$. AT2019avd has been interpreted as either an AGN flare or an unusual TDE. While the high Eddington ratio and MIR detection are similar to our observations of VT J1548, VT J1548 did was highly extincted and showed slower evolution in the MIR. Both of these difference could be caused by a larger dusty torus in VT1548 if it is undergoing the same type of flare as AT2019avd.

There is a growing population of transients which evolve on longer timescales than the typical TDE.  PS1-10adi \citep[][]{Kankare2017} was interpreted as a TDE candidate or highly obscured supernova in a Seyfert galaxy \citep[][]{Kankare2017}. This event was notable for its high bolometric luminosity (${\sim}10^{52}$ erg s$^{-1}$) and slow evolution: the optical light curve faded slowly over ${\sim}1000$ days after peaking at the Eddington luminosity. \cite{Kankare2017} proposed that it is a member of a class of similar transients; here we focus on PS1-10adi for simplicity. PS1-10adi also produced a dust echo, although the dust echo faded more quickly than that of VT J1548 and followed the expected blackbody temperature evolution. It was X-ray dim until ${\sim}1500$ days, at which point it brightened in the X-rays to ${\sim}10^{43}$ erg s$^{-1}$ and rebrightened briefly in the optical/IR. PS1-10adi was not detected in the radio at early times, but without further follow up we cannot exclude late time rebrightening. PS1-10adi also did not show strong coronal lines. Thus, VT J1548 and PS-10adi are similar in their high Eddington ratio, slow timescales, dust echoes, and late time X-ray detections, but there were clearly significant differences between this event and VT J1548. Some, but not all, of the differences can be explained if VT J1548 is observed on a more heavily obscured line of sight.

\tabcolsep=0.01cm
\begin{deluxetable*}{c |c|c|c|c|c|c|c|c|c|c|c|c|c}
\tablecaption{Comparison to select published transients  \label{tab:comp}}
\tablewidth{0pt}
\tablehead{\colhead{Name} & \colhead{$\log \frac{M_{\rm BH}}{M_\odot}$} & \colhead{BPT} & \colhead{$\frac{L_{\rm peak}}{L_{\rm edd.}}$} & \colhead{\makecell{Optically \\ dim?}} & \colhead{\makecell{Slow\\Evol.?}} & \colhead{MIR?} & \colhead{\makecell{Late time\\X-ray?}} & \colhead{Radio?} & \colhead{\makecell{Delayed\\radio?}} & \colhead{ECLE?} & \colhead{\makecell{Broad\\Balmer?}} & \colhead{\makecell{$E(B{-}V)_{\rm nuc.}$\\${\gtrsim} 0.3$?}} & \colhead{Trigger}}
\startdata \hline
AT2018dyk${}^1$ & $5.5$ & LINER & 0.004 & \xmark & \cmark & \xmark & \textbf{?} & \xmark & \textbf{?} & \cmark & \cmark & \xmark & AGN/TDE \\\hline
PS16dtm$^{2,3}$ & $6$ & NLSy1 & $2.8$ & \xmark & \cmark & \cmark & \textbf{?} & \xmark & \textbf{?} & \xmark & \cmark & \xmark & AGN/TDE \\ \hline
SDSS J1657+2345$^{4}$ & $6.2$ & AGN & $1.7$ & \cmark & \cmark & \cmark & \textbf{?} & \textbf{?} & \textbf{?} & \xmark & \cmark & \cmark & AGN/TDE \\\hline
AT2019avd$^{5,6}$ & $6.3$ & Comp. & $0.1$ & \xmark & \makecell{Double \\ peaked } & \cmark & \textbf{?} & \textbf{?} & \textbf{?} & \cmark & \cmark & \xmark & AGN/TDE \\ \hline
NGC 3599$^{7}$ & 6.4 & \makecell{Sey. 2 \\ LINER} & $0.004$ & \cmark & \cmark & \textbf{?} & \xmark & \textbf{?} & \textbf{?} & \xmark & \xmark & $-$ & AGN/TDE \\ \hline
\textcolor{red}{VT J1548} & \textcolor{red}{$6.48$} & \textcolor{red}{Comp.} & \textcolor{red}{0.1} & \textcolor{red}{\cmark} & \textcolor{red}{\cmark} & \textcolor{red}{\cmark} & \textcolor{red}{\cmark} & \textcolor{red}{\cmark} & \textcolor{red}{\cmark} & \textcolor{red}{\cmark} & \textcolor{red}{\cmark} & \textcolor{red}{\cmark} & \textcolor{red}{\textbf{AGN/TDE}} \\\hline
PS1-10adi$^{8,9}$ & $7$ & H\,II & ${\sim}1$ & \xmark & \cmark & \cmark & \cmark & \textbf{?} & \textbf{?} & \xmark & \cmark & \textbf{?} & TDE/SN \\\hline
ASASSN-15oi$^{10-12}$ & $7.1$ & H\,II & 0.15 & \xmark & \xmark & \xmark & \cmark & \cmark & \cmark & \xmark & \xmark & $-$ & TDE\\ \hline
1ES 1927+654$^{13}$ & $7.3$ & AGN & $0.01{-}0.2$ & \xmark & \xmark & \textbf{?} & \cmark & \textbf{?} & \textbf{?} & \xmark & \cmark & \cmark & AGN/TDE \\\hline
AT2017bgt$^{14}$ & $7.3$ & Comp. & ${\gtrsim}0.1$ & \xmark & \cmark & \textbf{?} & \cmark & \textbf{?} & \textbf{?} & \xmark & \cmark & \textbf{?} & AGN/TDE \\\hline
F01004-2237$^{15, 16}$ & $7.4$ & \makecell{H\,II \\ Sey. 2} & $0.02{-}0.7$ & \xmark & \cmark & \cmark & \textbf{?} & \xmark & \textbf{?} & \xmark & \xmark & $-$ & AGN/TDE \\\hline
OGLE17aaj$^{17}$ & $7.4$ & \textbf{?} & $0.01$ & \xmark & \cmark & \cmark & \textbf{?} & \xmark & \textbf{?} & \xmark & \cmark & \textbf{?} & AGN/TDE \\\hline
ASASSN-18jd$^{18}$ & $7.6$ & Comp. & $0.09$ & \xmark & \cmark & \cmark & \xmark & \textbf{?} & \textbf{?} & \cmark & \cmark & \xmark & AGN/TDE \\\hline
XMMSL2 J1446$^{19}$ & $7.8$ & H II & $0.02$ & \cmark & \cmark & \textbf{?} & \cmark & \xmark & \textbf{?} & \xmark & \xmark & $-$ & AGN/TDE \\ \hline
ASASSN-20hx$^{20}$ & $7.9$ & LLAGN? & 0.003 & \xmark & \cmark & \textbf{?} & \textbf{?} & \textbf{?} & \textbf{?} & \xmark & \xmark & $-$ & AGN/TDE \\ \hline
WISE J1052+1519$^{21}$ & 8.6 & AGN & $0.02$ & \xmark & \cmark & \cmark & \xmark & \textbf{?} & \textbf{?} & \xmark & \cmark & \xmark & \makecell{AGN \\ fading} \\ \hline
ASASSN-15lh$^{11,22,23}$ & 8.7 & LINER & $0.1$ & \xmark & \makecell{Double \\ peaked } & \cmark & \cmark & \xmark & \xmark & \xmark & \xmark & $-$ & SN/TDE \\ \hline
013815+00$^{24}$ & 9.3 & AGN & $0.02$ & \xmark & \cmark & \textbf{?} & \textbf{?} & \cmark & \xmark & \xmark & \makecell{$-$ \\ (broad Mg\,II)} & $-$ & AGN \\ \hline
\enddata
\tablecomments{See text for further description of select transients. Transients are sorted according to SMBH mass. SMBH masses are as reported by the authors, although we prefer to report those measured using the $M_{\rm BH}-\sigma_*$ relation. Late time detections refer to detections ${\sim}400$ days after the initial flare. Slow evolution refers to flares that rise over timescales $\gtrsim 50$ days or fade over a characteristic timescale ${\gtrsim}400$ days in the optical, IR, or X-ray. Eddington ratios are very approximate; they are reported using the peak bolometric luminosity when possible, otherwise using the peak luminosity in any given waveband. The trigger is as given in the relevant reference. Question marks refer to values for which we could not find a reported measurement. Note that WISE J1052+1519 is a fading CL AGN. References: ${}^{1}$\cite{Frederick2019}, ${}^{2}$\cite{Blanchard2017}, ${}^{3}$\cite{Jiang2017}, $^{4}$\cite{Yang2019}, ${}^{5}$\cite{Frederick2020}, ${}^{6}$\cite{Malyali2021}, $^{7}$\cite{Saxton2015}, ${}^{8}$\cite{Kankare2017}, ${}^{9}$\cite{Jiang2019}, ${}^{10}$\cite{Horesh2021}, ${}^{11}$\cite{Jiang2021_IRTDE}, ${}^{12}$\cite{Holoien2016}, ${}^{13}$\cite{Trakhtenbrot2019_1ES},  ${}^{14}$\cite{Trakhtenbrot2019}, ${}^{15}$\cite{Tadhunter2017}, ${}^{16}$\cite{Dou2017}, ${}^{17}$\cite{Gromadzki2019}, ${}^{18}$\cite{Neustadt2020}, $^{19}$\cite{Saxton2019}, ${}^{20}$\cite{Hinkle2021}, ${}^{21}$\cite{Stern2018}, ${}^{22}$\cite{Leloudas2016}, $^{23}$\cite{Margutti2017}, ${}^{24}$\cite{Kunert2020}. }
\end{deluxetable*}

PS1-10adi shows properties that are similar to the class of slowly-evolving flares reported by \cite{Trakhtenbrot2019}. That work focused on AT2017bgt, a slowly evolving (timescale $\gtrsim14$ months) optical/X-ray transient in an AGN (identified via archival X-ray detections) with $\log M_{\rm BH}/M_\odot \sim 7$. It showed strong, broad He\,II and Bowen fluorescence lines, as well as broad Balmer lines. \cite{Trakhtenbrot2019} proposed that this source, along with the similar transients OGLE17aaj and that hosted by the galaxy F01004-2237, form a new class of AGN flares where the UV/optical continuum emission increases by a factor ${\sim} 2$ in a few weeks. The flare in F01004-2237 was associated with a bright IR flare with ${\sim}$constant temperature that rose over thousands of days. Given the large number of similarities with VT J1548, it is feasible that VT J1548 is a member of this class but with a larger amount of dust and/or a more extincted sight line. Radio observations of AT2017bgt and like events are critical for assessing this interpretation.

Every transient discussed thus far has been detected in the optical. On the other hand, the candidate TDE or AGN flare SDSS J1657+2345 was discovered as a MIR flare with no optical counterpart \citep[][]{Yang2019}. It evolved over ${\sim}1000$ day timescales, like VT J1548. Broad H$\alpha$ is detected in its spectrum, but no broad H$\beta$ is detected. In contrast to VT J1548, no coronal line emission is detected. Follow up radio and X-ray observations would help determine whether this event is analogous to VT J1548. 

Similarly, none of the transients discussed have been reported to have unusual radio emission like that from VT J1548. While the radio luminosity of VT J1548 is typical of non-jetted TDEs \citep[][]{Alexander2020}, the SED and late time detections are atypical. We cannot exclude that most of the aformentioned transients show the same radio light curves as VT J1548: none of these transients have published, late-time, broadband radio follow up. If the late time emission is caused by an outflow colliding with a dense, torus-like medium, it is particularly important to obtain late-time radio follow up of transients where there is evidence for large obscuration.

The closest analog in the literature is the delayed radio emission from the TDE ASASSN-15oi reported by \cite{Horesh2021}. ASASSN-15oi rebrightened in the radio ${\sim}1400$ days after its initial flare. This event also rebrightened in the X-ray. The radio light curve evolved at an extremely fast rate, similar to VT J1548, and an inhomogeneous synchrotron model was required to fit the observations. Apart from this unusual radio emission, ASASSN-15oi was a relatively typical TDE, unlike VT J1548 \citep[][]{Holoien2016, Jiang2021_IRTDE}.

We conclude that VT J1548 is a unique transient, largely because of its large extinction, slow evolution, and delayed radio flare. While there is no single transient that definitively comes from the same class as VT J1548, by invoking different levels of obscuration it is plausible that the family of transients proposed by \cite{Trakhtenbrot2019} (AT2017bgt, OGLE17aaj, F01004-2237) and the IR transient SDSS J1657+2345 could form a class of similar objects.

\subsection{A qualitative model for VT J1548}

Next, we present a physical model that can explain all the above observations, and later we constrain the event that triggered VT J1548. In Figure~\ref{fig:cartoon}, we show a very qualitative cartoon model. At the center, we have shown an SMBH with an accretion disk. While we do not have direct evidence for an accretion disk, many of the scenarios we discuss in the rest of this section require a disk. Moreover, emission from an AGN-like disk and its corona could explain some of the observed X-rays. The typical outer radius of an AGN accretion disk is a few light days, or ${\sim}10^{-3}$ pc \citep[][]{Mudd2018}.

The clouds surrounding the accretion disk depict the broad line region, which produces the broad H$\alpha$. Given the width of the observed broad H$\alpha$, we expect that these clouds are located at a distance ${\sim}5\times10^{-3}$ pc. The BLR may have existed before the transient, as long as there was no significant ongoing accretion that would have illuminated the BLR and produced observable broad lines in the archival SDSS spectrum. Alternatively, the BLR could have formed via a dusty wind driven from the accretion disk, as has been proposed in some AGN models \citep[][]{Czerny2011}.

Outside of the BLR, we show coronal line-emitting clouds orbiting the SMBH, and a large dusty torus. The torus is not depicted as a standard doughnut, which is an oversimplification of the true structure, which fails to predict some observations \citep[e.g.][]{Mason2006, Ramos2009}. Instead, we adopt a clumpy, thick, flared, and extended gaseous disk. As discussed by \cite{Hopkins2012} and references therein, galactic-scale inflows trigger a series of gravitational instabilities on small scales, which produce a thick, eccentric disk near the SMBH without requiring active accretion. The orientation of the disk may be twisted and misaligned with the inner accretion disk, although we depict it as perfectly aligned for simplicity.

The observer is along a line of sight through the edge of the torus such that there is significant extinction, but the line of sight is not completely obscured as in Type 2 AGN. We expect the line of sight to have a column density $\log N_{\rm H}/{\rm cm}^{-2} \gtrsim 21.5$ given the constraints on the broad Balmer decrement.

We expect the torus to extend outward from at least ${\sim}0.4$ pc given the constraints from the MIR emission (Section~\ref{sec:IR}). At $0.4$ pc the temperature of the torus is ${\sim}1000$ K, and the dust interior to this radius is hotter. Dust that has been heated to $T_{\rm sub} \sim 1600$ K \citep[][]{Lu2016} is sublimated. 

The coronal line-emitting gas is represented by clouds at roughly the same distance (${\sim}0.8$ pc) as the MIR emitting gas and outflow. These clouds form from a hot, dusty wind driven by radiation pressure from the edge of the torus \citep[][]{Mullaney2008, Gelbord2009, Dorodnitsyn2012}. As the dusty clouds are accelerated off of the torus, the dust sublimates and releases the iron that produces the coronal line emission. While similar clouds are likely driven from the lower side of the torus, these would be highly extincted because they lie on a line of sight through the center of the torus. For simplicity, we do not draw them. While the geometry depicted may not produce the exact coronal line profiles observed, given uncertainties in the torus shape and dusty wind directions and kinematics, we are confident that there is a geometry which could replicate the observations.

Finally, we have drawn an outflow beginning at the accretion disk and that has collided with parts of the torus at a radius ${\sim}0.1$ pc, corresponding to the best-fit radius from our synchrotron model. This radius is roughly consistent with the distance to the MIR emitting dust. A wide angle outflow is required to explain the multiple synchrotron components \citep[see][for a discussion of possible origins]{Alexander2020}. We will discuss some of these possibilities in the following sections. While the exact position of this outflow is unknown, we emphasize that it need not be colliding with a uniform medium. Parts of the outflow may be incident on denser parts of the torus, and that could cause the unusual radio SED.

Coronal line emitters are generally interpreted as originating from one of three classes of transients: extreme AGN variability, tidal disruption events, or supernovae. In the following subsections, we discuss each of these possibilities in turn. We expect that our cartoon applies regardless of the exact cause of the flare, unless the flare was triggered by a slightly off-nuclear event (e.g., a supernova). In this case, we are observing the event through some abnormally thick cloud of material. We will discuss this possibility briefly in the following section.

\subsection{Is VT J1548 a supernova?}

We consider it unlikely that VT J1548 is caused by a supernova because of its luminosity and timescale. The difficulties of interpreting ECLEs as supernova have been discussed in many previous papers \citep[e.g.][]{Wang2011, Wang2012, Frederick2019}, so we only briefly consider it here. Only a few Type IIn supernova are observed to have coronal line emission. One of the SN IIn with the brightest coronal line emission was SN 2005ip, but by ${\sim}1000$ days the [Fe\,X] emission was only at ${\sim}10^{37}$ erg s$^{-1}$ \citep{Smith2009}, which is a factor of ${\sim}100$ dimmer than we observe. At no point during the evolution of SN 2005ip was the [Fe\,X] emission within a factor of ${\sim}10$ as bright as observed from VT J1548. Of course, VT J1548 may be the most extreme coronal line-emitting supernova seen to date. The X-ray luminosity ${\sim}10^{42-43}$ erg s$^{-1}$ required to produce the coronal lines is unprecedented for supernova $-$ one of the brightest, long-duration X-ray emitting supernova, SN1988Z, was only detected at ${\sim}10^{41}$ erg s$^{-1}$ \citep[][]{Fabian1996}.

The MIR emission from VT J1548 is difficult to reconcile with a supernova interpretation. Consider the case where the MIR photons are emitted by dust that is ejected by the supernova. The observed MIR emission is consistent with a distance ${\sim}0.4$ pc. To reach this radius within ${\sim}1$ year, the ejecta must have moved at a velocity ${\sim}c$. This is extraordinarily fast, so instead we invoke pre-existing material. The supernova either occurred in the galactic nucleus so that it is obscured by the torus, or the supernova is obscured by a torus-like quantity of dust outside the nucleus. Both of these scenarios are unusual, and combined with the extreme X-ray luminosity required to power the emission, we disfavor the supernova interpretation.

\subsection{Is VT J1548 a TDE?}

Next, we assess whether VT J1548 is consistent with a TDE. ECLEs are often attributed to TDEs \citep[e.g.][]{Wang2012}, although it is difficult to distinguish between AGN accretion variability (see next section) and TDEs. The observed coronal lines would be excited by the soft X-rays and UV continuum produced by the TDE \citep[e.g.][]{Wang2012}. A complication is that many TDEs show bright optical light curves \citep[e.g.][]{vanVelzen2021}, which we do not observe from VT J1548. However, an increasing number of optically-faint TDEs are being discovered \citep[see][for optically-faint X-ray selected TDEs]{Sazonov2021}. The optical emission from TC0221 may be heavily extincted (see example TDE lightcurves in Figure~\ref{fig:summary}). The flare may have occurred during a gap in survey coverage. Alternatively, the TDE may have been optically dim. TDEs associated with SMBHs with masses $\log M_{\rm BH}/M_\odot < 6$ may lack the optically thick gas layer which reprocesses higher energy photons and dominates the optical emission \citep{Lu2020}. 

The timescale of VT J1548 may also pose a problem: ``standard'' TDEs are expected to rise on short (${\sim}10$s of days) timescales, and they generally fade according to a canonical $t^{-5/3}$ power law \citep[see][for a review]{Gezari2021}. Example optical light curves are overplotted in Figure~\ref{fig:summary}. The IR emission from VT J1548 rises over ${\gtrsim}$2 years. As we discussed in Section~\ref{sec:IR}, a prompt, high-energy transient may be able to produce a slowly evolving, MIR flare. Because MIR photons emitted from the far side of the torus have to travel an extra distance ${\sim}2R_{\rm emit}$ for an emitting radius $R_{\rm emit}$, the flare is smoothed out over a time period $2R_{\rm emit}/c$.

If the observed MIR emission is the echo of a bright, prompt TDE, we have to invoke some delayed X-ray emission to explain our X-ray detections. We might expect dim, late-time X-ray detections from a viscous accretion disk, although whether such disks are expected is uncertain. \cite{vanVelzen2019} reported the detection of late-time (5-10 years post-flare) transient UV emission from eight optical TDE hosts which is inconsistent with this late-time models, but could be explained as emission from unobscured accretion disks with long viscous timescales. Similarly, \cite{Jonker2020} detected late time (5-10 years post-flare) X-ray emission from TDE candidates. Simulations of TDE evolution may have incorrectly predicted the late-time light curve evolution, possibly because of incorrect viscosity assumptions. If a slowly evolving viscous disk is present in SDSS J1548, we would expect the MIR flare to fade extremely slowly (i.e., decades timescale).

Late-time interactions between an outflow launched during the initial TDE and a dusty torus are also able to produced delayed X-ray emission at a luminosity ${\sim}10^{41-42}$ erg s$^{-1}$ \citep[][]{Mou2021}. This model can also explain the brightening in the radio via shocks due to the outflow hitting the torus, and predicts that this event should be $\gamma$-ray bright \citep[][]{Mou2021B}.

Alternatively, we may be witnessing a TDE that evolves slowly because of delayed accretion disk formation \citep[although see][]{vanVelzen2019, Jonker2020}. TDE accretion disks may form when stellar debris streams collide because of general relativistic precession, eventually dissipating enough energy to collapse \citep[][]{Guillochon2015}. The precession is correlated with the SMBH mass: stellar streams orbiting SMBHs with $\log M_{\rm BH}/M_\odot \lesssim 6$ may take years for the debris to precess sufficiently to cause collisions \citep[][]{Guillochon2015}. The slow disk formation erases information about the mass fallback rate which usually sets the light curve decay time to $t^{-5/3}$. TDEs with delayed accretion disks decay following a power law ${\sim}t^{-1}$ \citep[][]{Guillochon2015}.

This delayed accretion disk model also requires no pre-existing accretion disk. We have invoked a torus to explain the IR emission from VT J1548, but some models predict that tori are only hosted by AGN with sufficiently large luminosities (${\gtrsim}10^{39}$ erg s$^{-1}$ for a $\log M_{\rm BH}/M_\odot \sim 6$ SMBH; \citealp[][]{Honig2007}). As discussed by \cite{Hopkins2012}, it is feasible that tori can form in quiescent galaxies if dynamical instabilities reminiscent of the bars-within-bars models drive gas to the galactic center. Regardless, we consider the possibility that SDSS J1548 had a pre-existing accretion disk for completeness.

Like TDEs in quiescent galaxies, it is feasible that TDEs in AGN may produce emission years after the initial flare. However, predictions for the observational characteristics of TDEs in AGN are limited. \cite{Chan2020} modeled a TDE in an AGN and predicted light curves that evolve on ${\sim}$month timescales, which is much faster than observed for VT J1548. However, the simulations spanned a very small range of parameter space, so we will have to wait for a more expansive set of models of TDEs in AGN to constrain whether that mechanism could have triggered VT J1548.

In summary, interpreting VT J1548 as a TDE is plausible. In one scenario, the torus geometry and optical depth are such that the long duration MIR brightening can be produced by a short high energy flare. Alternatively, TDEs in AGN may simply be able to evolve on very long timescales.

\subsection{Is VT J1548 an AGN flare?} \label{sec:agn}

VT J1548 may be an extreme AGN event if it is neither a TDE nor a supernova. Variable obscuration can cause bright flares on the orbital timescale of the obscuring cloud \citep[e.g.][]{Stern2018} but could not explain most all of the other transient emission, such as the broad H$\alpha$, and we do not consider it further. 

Alternatively, VT J1548 may be a changing look AGN (CL AGN). CL AGN comprise a rapidly growing class of AGN which transition from Type 1, with a weak continuum and narrow line emission, to Type 1.8-2, with broad H$\alpha$ and H$\beta$ alongside a strong continuum, or vice versa. These transitions occur on year-decade timescales, and the AGN can fade and/or rise \citep[e.g.][]{Sheng2017}. They are sometimes associated with MIR \citep[][]{Sheng2017, Stern2018, Sheng2020} and X-ray \citep[][]{Parker2016} variability and flaring. 

If VT J1548 is a CL AGN, it is unprecedented. Unlike most CL AGN, SDSS J1548 showed no strong pre-flare variability or evidence for AGN activity \citep[e.g., see candidates in][]{Yang2018}, although this might be a selection effect since CL AGN candidates are often identified by their pre-flare variability. No unambiguous CL AGN with the unusual radio emission and high Eddington ratio of VT J1548 has been observed to date, although radio follow up of CL AGN is limited. The CL AGN 013815+00 is notable because it brightens in both the optical and the radio \citep[][]{Hameury2020}. Similarly, Mrk 590  underwent multiple transitions between Seyfert types in the last thirty years and shows some evidence for radio variability \citep[][]{Koay2016, Yang2021}. A more complete sample of radio-selected, optical/UV/IR flaring CL AGN, or a comprehensive follow up program to measure the radio light curves of ongoing CL AGN are key to understanding the expected radio signatures.

Flaring galaxies with $\log M_{\rm BH}/M_\odot \lesssim 8$ are difficult to unambiguously classify as a TDE or AGN flare, in part because AGN accretion disks still present mysteries: observations of large amplitude variability in AGN are becoming common, and most accretion disk models do not predict frequent, large variability, but instead explain these flares by ``instabilities'' \citep[][]{Lawrence2018}. Large uncertainties in the relevant timescales and flare amplitudes renders any comparison to observation difficult.

NGC 3599 exemplifies this difficulty. This galaxy underwent a slow, soft X-ray flare that rose over multiple years \citep[][]{Saxton2015}. It lacked observations between the rise and decay, which complicated the interpretation. However, the slow timescale is atypical of normal, prompt TDEs, although, as we have discussed, it is likely that some TDEs can evolve on much longer timescales and cannot be excluded as a trigger of the flare in NGC 3599. \cite{Saxton2015} suggested the Lightman-Eardley disk instability as one possible cause of the flare. The instability arises because \cite{Shakura1973} thin accretion disks become unstable when radiation pressure dominates over thermal pressure. This condition is fulfilled in the inner regions of any disk that is accreting at a near-Eddington luminosity. The instability manifests as a limit-cycle behavior. When the disk is bright and highly accreting, the inner disk is unstable and empties, which reduces the accretion rate. The inner disk slowly refills, eventually returning to the high accretion state and repeating the cycle. The rise time to the high accretion rate state is set by the time required to heat the inner disk, which depends strongly on the viscosity prescription but must be greater than \citep[][]{Saxton2015}
\begin{equation}
    R_{\rm trunc}/c_s \sim 1\,{\rm month} \bigg( \frac{R_{\rm trunc}}{100 R_g} \bigg) \bigg( \frac{M_{\rm BH}}{10^6\,M_\odot} \bigg)
\end{equation} where $R_{\rm trunc}$ is the radius at which the disk will become truncated and $c_s$ is the sound speed. $R_g$ is the gravitational radius of the SMBH. After the rise, the emission plateaus for an unconstrained time as the inner disk is cleared out. Once the inner disk is empty, the emission will decay to the low state. The decay time is poorly constrained, but it is expected to be faster than the rise time.

Like NGC 3599, VT J1548 evolves on the correct timescales to be explained by the Lightman-Eardely instability. In the future, we can more definitively constrain this possibility by monitoring the evolution of VT J1548 for evidence of (1) a decay time that is much more rapid than the rise time and (2) a repeat flare on a many-decade timescale. Even if such behavior is observed, the interpretation is complicated. The Lightman Eardley instability cannot be considered in isolation: other instabilities are predicted in the inner disk. For example, the ionization instability applies to cool ($T\sim6000$ K) regions of the disk and is the result of the strong temperature and density dependence of the opacity of partially ionized hydrogen \citep[][]{Hameury2020}. 

Moreover, it is unclear whether the Lightman Eardley instability actually occurs in low mass accreting SMBHs, let alone AGN \citep[][]{Janiuk2011}. The \cite{Shakura1973} viscosity is an oversimplification, and there is some evidence that a more physical viscosity prescription eliminates the instability \citep[][]{Blaes2006}. Similarly, while the ionization instability is well established for dwarf novae \citep[see][for a review]{Hameury2020}, it has not been definitively observed in AGN.

In summary, VT J1548 may be a CL AGN, although its lack of strong AGN signatures and bright radio flare are unusual. Observations of a repeat flare or a fast decay time could support a CL AGN origin.

\subsection{Future work and observations}

It is extremely difficult to unambiguously determine the cause of VT J1548, as is a common issue for like transients. In many cases, extreme AGN variability is as feasible an explanation as a TDE-like transient. In this section, we suggest future observations and theoretical work that could help constrain the origin of VT J1548 and like events. We begin with possible observations.
\begin{itemize}
    \item {\it Early time spectroscopic follow up of VT J1548-like transients.} While inapplicable to VT J1548 itself, this follow up could help constrain the presence of features such as He\,II and Bowen flourescence lines that may have faded by the late time observations of VT J1548.
    \item {\it Long term NIR/MIR monitoring of VT J1548 (and similar transients).} Knowledge of the IR evolution is essential to constrain the origin of the flare. If this event is triggered by a TDE, we expect the emission to begin fading soon. If VT J1548 is an AGN flare, it could remain bright for decades or longer. Combined with theoretical modelling of AGN flares and TDE in AGN, the fade time of the event may constrain its origin. Some AGN flares may fade more quickly than a typical TDE \citep[][]{Saxton2015}. We are actively monitoring VT J1548 in the NIR. If SDSS J1548 is also monitored in the MIR for multiple decades, we could also constrain the presence of a repeat flare, which may provide a smoking gun for extreme AGN variability.
    \item {\it Long term optical spectroscopic monitoring.} We are actively following up VT J1548 with optical spectrographs to determine the evolution of the broad Balmer emission, the coronal lines (both profile and flux), as well as any other features that may begin to evolve. 
    \item {\it Long term X-ray/Radio (100s of MHz $-$ GHz) monitoring.} Long term X-ray/radio monitoring will allow us to constrain the origin of emission at both wavelengths. As we have mentioned the importance of this follow up throughout the text, we do not discuss it further.
    \item {\it Optical IFU follow up}. Optical IFU observations would allow us to constrain whether the pre-existing high ionization emission, such as the [O\,III] lines, are nuclear or very extended. Then, we could constrain the history of AGN activity.
\end{itemize}
This list is far from exhaustive (e.g., polarimetric observations and hard X-ray spectra would prove useful). 

On the theoretical side, the most critical work is extensive simulations of TDEs in AGN-like environments. In particular, given the difficulty of getting early-time follow up of these events, detailed simulations of the fading of the TDE emission would be valuable. It would also be useful to model the response of an AGN torus to a TDE-like flare accounting for different torus models, inclination angles, and flare durations/shapes. Finally, detailed models of the expected evolution of coronal line flux and profile during a TDE-like event would prove extremely valuable towards constraining the timescales of ECLEs, and hence their triggers.

\section{Conclusions} \label{sec:conclusion}

We have presented the first radio selected ECLE, VT J1548, and its host, SDSS J1548. This work can be summarized as follows:
\begin{enumerate}
    \item VT J1548 is associated with a MIR flare that rose in ${\sim}900$ days and has plateaued with a constant color corresponding to a blackbody with $T\sim1000$ K and $L\sim 0.1 L_{\rm edd.}$ (Figure~\ref{fig:summary}). Radio emission turned on during the WISE flare. The radio SED can be modeled as synchrotron from an outflow incident on an imhomogeneous medium. No optical flare is detected. Transient, X-ray emission with $L_{0.2-10\,{\rm keV}} \sim 10^{42}$ erg s$^{-1}$ was detected ${\sim}1000$ days after the MIR flare began.
    \item Transient coronal lines with $L_{\rm cor.} \sim L_{\rm [O\,III]}$ and broad H$\alpha$ emission are detected ${\sim}1000$ days post-flare. The coronal line emission from VT J1548 is double peaked with a velocity separation ${\sim}230$ km s$^{-1}$ (Figure~\ref{fig:profiles}). The broad H$\alpha$ emission has FWHM $\sim 1900$ km s$^{-1}$. No broad H$\beta$ is detected, suggesting a high extinction $E(B{-}V) \gtrsim 0.7$ (Figure~\ref{fig:broad}). 
    \item SDSS J1548, the host of VT J1548, is an S0 galaxy at $z=0.031$ ($d_L = 137$ Mpc). Its pre-flare line ratios are consistent with no or weak AGN activity (Figure~\ref{fig:BPT}). Its pre-flare WISE color was inconsistent with typical AGN colors and it showed no significant MIR variability. It hosts a low mass SMBH with $\log M_{\rm BH}/M_\odot = 6.48 \pm 0.33$. 
    \item VT J1548 is unique when compared to other transients, although it shares individual properties with other objects (Table~\ref{tab:comp}). It is reminiscent of the class reported by \cite{Trakhtenbrot2019} and the MIR flare SDSS J1657+2345 \citep[][]{Yang2019}, which are all slowly evolving. None of these events emit in the radio like VT J1548, nor do they show coronal line emission. The only transient with radio emission which resembles that from VT J1548 is the TDE candidate ASASSN-15oi, which was detected at late-times in the X-ray and radio with an unusual radio SED, although none of its other properties resemble VT J1548.
    \item VT J1548 can be modeled with a broad line region (${\sim}10^{-3}$ pc) surrounded by a dusty torus (Figure~\ref{fig:cartoon}). At the inner edge of the torus (${\sim}0.4$ pc), sublimated dust accelerated in a radiation-driven wind causes the formation of coronal line-emitting clouds. Alternatively, the coronal line-emitting clouds are orbiting the SMBH. Some of the dust in the torus is heated to produce the MIR flare. The synchrotron-emitting outflow is launched near the SMBH and collides with the torus.
    \item VT J1548 could plausibly have been triggered by a tidal disruption event or an AGN disk instability. In the TDE scenario, the high observed Eddington ratio, the radio emission, and the slow evolution are unusual. In the AGN scenario, the lack of strong pre-flare AGN activity is uncommon, the high Eddington ratio is unexpected, and the unusual radio emission is unprecedented.
\end{enumerate}

We have emphasized the difficulty of distinguishing between an AGN flare and a TDE in a highly obscured galaxy with evidence for weak AGN activity. Such efforts are particularly complicated because of the freedom in structure and optical depth of a torus, which can eliminate most all timescale information if the inner ${\sim}$pc of the galaxy is too extincted to be directly visible. Moreover, the uncertainty in the timescale, luminosity, and multiwavelength properties of flares from TDEs (in both AGN and quiescent galaxies) as well as AGN disk instability-driven flares renders it difficult to distinguish between these events even with early time follow up. While it may be difficult to constrain the origins of individual events, population studies are key to characterizing the range and relative frequency of these different flares. The identification of {\it classes} of transients is likely to be less ambiguous if members are observed from a range of inclination angles over a variety of time baselines in multiple wavebands.

\acknowledgements

We would like to thank Sterl Phinney for useful discussions. We would like to thank Phil Hopkins for insight into the formation and evolution of AGN torii. Finally, we would like to thank the entire ZTF TDE/AGN group for useful discussions, and in particular Suvi Gezari and Sjoert van Velzen. 

CJL acknowledges support from the National Science Foundation under Grant No.\ 2022546.

This work made use of data supplied by the UK Swift Science Data Centre at the University of Leicester. The National Radio Astronomy Observatory is a facility of the National Science Foundation operated under cooperative agreement by Associated Universities, Inc. CIRADA is funded by a grant from the Canada Foundation for Innovation 2017 Innovation Fund (Project 35999), as well as by the Provinces of Ontario, British Columbia, Alberta, Manitoba and Quebec. Some of the data presented herein were obtained at the W. M. Keck Observatory, which is operated as a scientific partnership among the California Institute of Technology, the University of California and the National Aeronautics and Space Administration. The Observatory was made possible by the generous financial support of the W. M. Keck Foundation. The authors wish to recognize and acknowledge the very significant cultural role and reverence that the summit of Maunakea has always had within the indigenous Hawaiian community.  We are most fortunate to have the opportunity to conduct observations from this mountain. Funding for the SDSS and SDSS-II has been provided by the Alfred P. Sloan Foundation, the Participating Institutions, the National Science Foundation, the U.S. Department of Energy, the National Aeronautics and Space Administration, the Japanese Monbukagakusho, the Max Planck Society, and the Higher Education Funding Council for England. The SDSS Web Site is \url{http://www.sdss.org/}. The SDSS is managed by the Astrophysical Research Consortium for the Participating Institutions. The Participating Institutions are the American Museum of Natural History, Astrophysical Institute Potsdam, University of Basel, University of Cambridge, Case Western Reserve University, University of Chicago, Drexel University, Fermilab, the Institute for Advanced Study, the Japan Participation Group, Johns Hopkins University, the Joint Institute for Nuclear Astrophysics, the Kavli Institute for Particle Astrophysics and Cosmology, the Korean Scientist Group, the Chinese Academy of Sciences (LAMOST), Los Alamos National Laboratory, the Max-Planck-Institute for Astronomy (MPIA), the Max-Planck-Institute for Astrophysics (MPA), New Mexico State University, Ohio State University, University of Pittsburgh, University of Portsmouth, Princeton University, the United States Naval Observatory, and the University of Washington. Based on observations obtained with the Samuel Oschin 48-inch Telescope at the Palomar Observatory as part of the Zwicky Transient Facility project. ZTF is supported by the National Science Foundation under Grant No. AST-1440341 and a collaboration including Caltech, IPAC, the Weizmann Institute for Science, the Oskar Klein Center at Stockholm University, the University of Maryland, the University of Washington, Deutsches Elektronen-Synchrotron and Humboldt University, Los Alamos National Laboratories, the TANGO Consortium of Taiwan, the University of Wisconsin at Milwaukee, and Lawrence Berkeley National Laboratories. Operations are conducted by COO, IPAC, and UW.

\facilities{Swift(XRT), Keck I(LRIS), Keck II(ESI), VLA, WISE, XMM-Newton(EPIC)}
\software{Astropy \citep{Robitaille2013Astropy:Astronomy, Astropy-Collaboration:2018},
          Matplotlib \citep{Hunter2007Matplotlib:Environment},
          NumPy \citep{numpy, 2011CSE....13b..22V},
          Pandas \citep{reback2020pandas, mckinney-proc-scipy-2010},
          SciPy \citep{2020SciPy-NMeth},
          dynesty \citep{2020MNRAS.493.3132S, 2004AIPC..735..395S, 2019S&C....29..891H},
          emcee \citep{Foreman2013},
          pybdsf \citep{Mohan2015},
          }

\appendix

\section{VLASS Transient Search} \label{sec:VLASStrans}

We identified transient sources between the VLASS Epoch 2.1 and Epoch 1 observations using the following procedure. First, we run the source extractor \texttt{PyBDSF} \citep[][]{Mohan2015} on the VLASS Epoch 2.1 quicklook images provided by the National Radio Astronomy Observatory (NRAO) \citep[][]{Lacy2020}, spanning ${\sim}17,000$ deg$^2$. We identify point source candidates as regions within the Epoch 2.1 images where contiguous ``islands" of $> 4\sigma$ pixels surrounding a peak pixel of $> 6 \sigma$ can be well described by a single 2D Gaussian. Some of these candidates are due to deconvolution artifacts: typically sidelobes near bright sources or extended stripes. We flag the majority of these artifacts in an automated way using a stripe detection algorithm and by comparing the pixels near the source to the pixels in the 1 arcminute region around it (Dong et al. in prep). After flagging likely artifacts, we estimate the flux of each point source candidate as its peak pixel value, and the uncertainty as the local value in the quicklook RMS maps provided by the NRAO. We then check the corresponding location in the Epoch 1.1 image data. Based on the local pixel values in each epoch, we estimate the probability of variability by comparison with a grid of Monte Carlo simulations of sources embedded in Gaussian noise (Dong et al. in prep). We create an initial transient catalog in which we retain sources that have (1) no artifact flags, (2) a $>$90\% probability of being variable, (3) a peak $>$7$\sigma$ in Epoch 2.1, and (4) a peak $<$ 3$\sigma$ in Epoch 1.1. We visually inspected all transient candidates in the initial catalog, removing the artifacts that were missed by our automated filters. The remaining sources comprise our final transient catalog.

\section{Spectral Fitting Methods} \label{sec:spectral_fitting}

We use a consistent method to fit the optical emission lines in all of our observations. First, we correct the spectrum for Milky Way extinction using $A_V=0.1606$ and $R_V=3.1$ \citep{Schlafly2011, Fitzpatrick1999}. We remove the stellar continuum using a full spectrum fit with the penalized pixel-fitting (\texttt{pPXF}) method. We use the implementation of \texttt{pPXF} from \cite{Cappellari2004, Cappellari2017} with the default MILES templates \citep[][]{Vazdekis2010}. We run the fit using recommended procedures to determine the appropriate regularization error and refer the reader to the \texttt{pPXF} documentation for details. Rather than mask the emission lines during this fit, we include Gaussian components for each emission line, and allow the parameters for narrow forbidden, narrow allowed, and broad lines to float separately. These Gaussian fits are not used to measure the line fluxes; we only include emission line components to prevent the lines from biasing the \texttt{pPXF} fit. Our exact treatment of the emission lines does not affect our results. We also include a multiplicative normalization component that is a degree $10$ polynomial.

The MILES templates cover the wavelength range $3525-7500\,{\rm \AA}$, which does not span the full wavelength range of our observations. Hence, we perform an additional median-subtraction to normalize the remaining parts of the spectrum, as well as to correct for any continuum flux that was poorly removed by the \texttt{pPXF} fit. First, we subtract the best-fit stellar continuum and emission lines found by \texttt{pPXF} from our observed spectrum. We subtract the value of the best-fit stellar continuum at the nearest available pixel from the portions of the spectrum not covered by the template fit. From this procedure, we have a preliminary continuum-subtracted spectrum. Next, we median smooth this preliminary continuum-subtracted spectrum with a kernel of ${\sim}130$ pixels, which we found was sufficient for both the LRIS and SDSS spectra. We identify all points in the spectrum that are $>10\sigma$ from the resulting median smoothed continuum and mask them, as well as the ${\sim}10$ pixels neighboring those points. Then, we median smooth the preliminary continuum-subtracted spectrum a second time with these pixels masked. This gives us a correction to the continuum, which we subtract from our preliminary continuum-subtracted spectrum to obtain the final continuum-subtracted spectrum. 

Next, we measure the emission line fluxes from the final continuum-subtracted spectrum. We model each line with as many Gaussian profiles are required, and we specify throughout the text any case where multiple components are needed. We let the width of the Gaussians float independently for different lines, and include a broad component if necessary. We also include a linear continuum component to account for any residual flux. We fit each emission line separately unless multiple emission lines are so close that they cannot be fit independently (e.g., the H$\alpha$ and [N\,II] lines). We fit a region around each line that includes a ${\sim}10-20\,{\rm \AA}$ continuum region.

We run the emission line fit using the \texttt{dynesty} Nested Sampler \citep[][]{Speagle2020, Skilling2004}. Our stopping condition is $\Delta \log \mathcal{Z} = 5$, which we verify does not affect our results. Unless otherwise specified, we report $1\sigma$ errors on all line fluxes.

\section{Measurement of the Bulge Velocity Dispersion} \label{sec:esi}

We measure the bulge velocity dispersion using the $R\sim13,000$ ESI spectrum of SDSS J1548 following the methodology in \cite{Wevers2017}. To ensure that we are measuring the \textit{bulge} velocity dispersion, we consider two different methods of extracting the host spectrum: first, we use the spectrum extracted from the full slit; second, we isolate the bulge by using the spectrum extracted from a region centered on the peak galaxy light and with width $0\farcs5$, which roughly corresponds to the seeing during the observation. We find no significant difference in the results.

We use the \cite{Cappellari2017} implementation of \texttt{pPXF} to measure the velocity dispersion. First, we mask all emission lines in the spectrum, including the Hydrogen Balmer lines and all TDE features. We fit the spectrum using the \cite{Prugniel2007, Prugniel2001} high resolution ($R\sim 10000$) template library. We run the fit using the recommended \texttt{pPXF} settings. The best fit velocity dispersion is ${\sim}66$ km s$^{-1}$. Removing the intrinsic resolution (${\sim}22$ km s$^{-1}$) only changes the result by a few km s$^{-1}$. Next, we convert to SMBH mass using the relation from \cite{Ferrarese2005}. The intrinsic scatter in the $M_{\rm BH}-\sigma_*$ relation strongly dominates our results, although we also propagate through both the uncertainties in the velocity dispersion measurement and assumed uncertainties in the intrinsic resolution. We find an SMBH mass $\log M_{\rm BH}/M_\odot = 5.98 \pm 0.38$. Alternatively, the more recent $M_{\rm BH}-\sigma_*$ relation from \cite{Kormendy2013} gives $\log M_{\rm BH}/M_\odot = 6.48 \pm 0.33$, which is consistent with the \cite{Ferrarese2005} within $1\sigma$. We adopt the latter calibration because it includes more low mass galaxies.

\bibliography{main.bib}{}
\bibliographystyle{aasjournal}
\end{document}